\documentclass[reprint, preprintnumbers, aps, prd, floatfix]{revtex4-2}

\usepackage{amssymb,amsmath,physics,graphicx,xcolor,hyperref}

\newcommand{\bham}{Institute for Gravitational Wave Astronomy \& School of Physics and Astronomy, University of Birmingham, Edgbaston, Birmingham B15 2TT, UK}

\begin{document}

\title{Modelling the Ringdown from Precessing Black Hole Binaries}

\author{Eliot Finch}
\email{efinch@star.sr.bham.ac.uk}
\affiliation{\bham}

\author{Christopher J. Moore}
\email{cmoore@star.sr.bham.ac.uk}
\affiliation{\bham}

\date{\today}

\begin{abstract}
    Modelling the end point of binary black hole mergers is a cornerstone of modern gravitational-wave astronomy.
    Extracting multiple quasinormal mode frequencies from the ringdown signal allows the remnant black hole to be studied in unprecedented detail.
    Previous studies on numerical relativity simulations of aligned-spin binaries have found that it is possible to start the ringdown analysis much earlier than previously thought if overtones (and possibly mirror modes) are included. 
    This increases the signal-to-noise ratio in the ringdown making identification of subdominant modes easier.
    In this paper we study, for the first time, black hole binaries with misaligned spins and find a much greater variation in the performance of ringdown fits than in the aligned-spin case. 
    The inclusion of mirror modes and higher harmonics, along with overtones, improves the reliability of ringdown fits with an early start time; 
    however, there remain cases with poor performing fits.
    While using overtones in conjunction with an early ringdown start time is an enticing possibility, it is necessary to proceed with caution. 
    We also consider for the first time the use of numerical relativity surrogate models in this type of quasinormal mode study and address important questions of accuracy in the underlying numerical waveforms used for the fit.
\end{abstract}

\maketitle

\section{Introduction}

The gravitational wave (GW) observatories LIGO \cite{2015CQGra..32g4001L} and Virgo \cite{2015CQGra..32b4001A} have now observed dozens of GW events \cite{2019PhRvX...9c1040A, 2020arXiv201014527A}, mostly from binary black hole (BBH) mergers. Particularly prominent in the GW signals of the higher-mass systems are the final few wave cycles, known as the \emph{ringdown}, emitted as the system settles into its final state: a Kerr black hole (BH). The ringdown signal contains a superposition of oscillatory modes, the frequency spectrum of which is characteristic of the remnant BH.

The characteristic oscillations of the remnant BH are called \emph{quasinormal modes} (QNMs), so-called because, unlike normal modes, they decay over time. The QNM frequencies are complex, $\omega = 2\pi f - {i}/\tau$, with the real part $f$ giving the oscillation frequency and the reciprocal of the imaginary part $\tau$ giving the damping time. The QNM frequencies can be calculated within the framework of linearised gravity, treating the gravitational field in the vicinity of the remnant as a small (linear) perturbation of the Kerr metric \cite{qnms}.
Therefore, the QNM description of the GW signal is only expected to be valid at sufficiently late times, when the nonlinearities from the merger have largely decayed away. 

The remnant Kerr BH has no hair; it is fully described by only a final mass, $M_f$, and a dimensionless final spin parameter, $\chi_f = |\vb*{\chi}_f|$. The same is true of the spectrum of QNM frequencies, $\omega_{\ell m n}(M_f, \chi_f)$, which are also functions of only the mass and spin. Individual QNMs are indexed by the triplet $(\ell, m, n)$ which are the polar ($\ell\geq2$), azimuthal ($-\ell \leq m \leq \ell$) and overtone ($n \geq 0$) numbers respectively. 
The spectrum is further complicated by the fact that QNMs occur in pairs. A complete description of the ringdown must include the \emph{mirror modes} $\omega'_{\ell m n}$ \cite{qnms, bcw, mirror_modes, 2014PhRvD..90l4032L} with negative real frequency $f'_{\ell m n}$ along with the \emph{regular modes} $\omega_{\ell m n}$ with $f_{\ell m n}>0$
\footnote{We choose to classify QNMs as either \emph{regular} or \emph{mirror}. This is closely related to, but still distinct from, the prograde/retrograde classification of QNMs used in, for example, \cite{gwtc-2_tests}.}.
A quantification of the mirror modes was treated in Appendix D of
\cite{2020PhRvD.102d4053J}; some of these estimates were later
confirmed in \cite{mirror_modes}.
The spectrum of mirror modes contains the same information as the regular modes (albeit with nontrivial relationships between them, see Eqs.~\ref{eq:sym_mirror_modes_conj}) which has sometimes led to them being neglected. Whether they can, in fact, be neglected will depend on the relative excitation amplitudes of the regular and mirror modes and their differing decay times. In general, the ringdown will contain a superposition of all these modes with different excitation amplitudes and phases (see Eq.~\ref{general_ringdown}). Usually, the GW strain is dominated by the $\ell=|m|=2$ modes. Furthermore, the overtones decay more quickly (i.e.\ $\tau$ decreases) with increasing $n$ so that at late times the signal will be dominated by the fundamental $n=0$ modes. Therefore, the most prominent QNM in the ringdown is expected to be the $(\ell, m, n)=(2,2,0)$ mode, and the observational challenge is usually to detect the presence of other, subdominant modes.

The study of QNMs has applications in both astro and fundamental physics. The highly constrained dependence of the QNM spectrum on only the remnant mass and spin means that, conversely, if a QNM frequency is measured, then the mass and spin of the final BH merger can be inferred. For high-mass systems, where only the ringdown signal is observable, this may be the only information available about the nature of the source \cite{bcw, 2020PhRvD.101h4053B}. For lower-mass systems, measuring QNM frequencies allows us to estimate the remnant properties independently of the rest of the signal, and so consistency tests can be performed. 
For example, a test of the BH area theorem can be performed in this way \cite{2018PhRvD..97l4069C, 2020arXiv201204486I}. A similar consistency test using full inspiral-merger-ringdown models and a sharp cut in the frequency (rather than time) domain was performed on GW150914 \cite{tgr}. Each additional QNM that can be detected in the ringdown provides a separate estimate of the mass and spin of the remnant. Therefore, if multiple QNM frequencies can be identified, a ringdown-only consistency test on the expected Kerr-like nature of the remnant BH can be performed \cite{2004CQGra..21..787D, 2019PhRvD..99l3029C} (this is possible only if the $(\ell, m, n)$ of the modes are known). In these tests, deviations from the expected results may point to new physics beyond general relativity. 

QNMs also have practical uses in waveform modelling.
They are used in full inspiral-merger-ringdown BBH waveforms produced in both the phenomenological \cite{2020arXiv200406503P, 2020PhRvD.102f4002G, 2020PhRvD.102f4001P} and effective-one-body approaches \cite{2007PhRvD..75l4018B, 2007PhRvD..76j4049B, 2011PhRvD..84l4052P}.

A prerequisite for any ringdown analysis is a suitable choice for the \emph{start time}, $t_0$, of the ringdown. Starting too early risks a GW signal contaminated with nonlinearities that cannot be described by a model based solely on QNMs and obtaining biased measurements as a result. On the other hand, starting too late leaves a short signal that is already decaying and without enough signal-to-noise to make useful measurements. The difficulties of defining a suitable start time and some surprising data analysis consequences of this were discussed in \cite{2017arXiv170605152T}. Previous studies on numerical relativity (NR) waveforms have tended to use just a single \cite{1998PhRvD..57.4535F} or relatively small number ($\leq 4$) \cite{2012PhRvD..85b4018K, 2014PhRvD..90l4032L, 2018PhRvD..98j4020C, 2018PhRvD..97j4065B} of QNMs, and have used a wide range of start times. Typically, the start time is referred to the maxima of some time-dependent quantity (e.g.\ the 22 mode of the strain, the modulus of the Weyl scalar, or the total GW luminosity; these quantities peak at times that typically differ by a few tens of $M$) and typical choices were ${10-20M}$ after the peak (although, see \cite{2007PhRvD..76f4034B} where a range of different start times are explored). Several studies caution against starting too early \cite{2020PhRvD.102d4053J, 2020PhRvD.101d4033B}.

More recently, \cite{overtones} (see also \cite{2020PhRvD.101j4005O}) looked to allow the ringdown to start significantly earlier by including multiple overtones. In particular, \cite{overtones} found that by including up to seven overtones the ringdown analysis can be started as early as the peak in the 22 mode of the strain. This might be considered a surprising result; the signal peak is expected to occur when the remnant BH (to the extent that it yet even makes sense to consider it as such) is most highly distorted and linear perturbation theory is not expected to be valid. The failure of this intuition was investigated in \cite{2020arXiv200400671O} which suggests much of the nonlinearity is trapped behind a forming common apparent horizon and never makes it out to future null infinity in the form of GWs. Even more surprising, in \cite{mirror_modes} this approach was extended by the inclusion of mirror modes along with overtones (thereby doubling the number of QNMs) and it was found that it was possible to start the ringdown analysis even earlier (up to ${10M}$ \emph{before} the peak). Clearly it is not surprising that a model with so many free parameters is able to fit the GW signal well; the important point is that it is able to do so without obtaining biased values for the final mass and spin. A ringdown model with overtones was successfully applied to GW150914 in \cite{tnh} to extract the fundamental QNM and the first overtone from the noisy data, and more recently to events in GWTC-2 \cite{gwtc-2_tests}. 

To the best of our knowledge, previous studies modelling the ringdown from NR simulations with QNMs have only considered aligned-spin BBH systems. We note that some work on precessing systems has been done in the extreme mass ratio limit, see \cite{2019PhRvL.123p1101H,2019PhRvD.100h4032L}. It is well known that misalignment between the orbital angular momentum and the spins of the component BHs cause the orbit to precess during the inspiral phase of the evolution leading to qualitatively different GW signals at early times (e.g.\ see \cite{1994PhRvD..49.6274A}). It is less clear what effect, if any, misaligned component spins would have on the late time ringdown signal which is generally associated with the remnant BH. The primary aim of this paper is to address this question by systematically extending the analyses of \cite{overtones, mirror_modes} to a large number of precessing BBH simulations from the SXS catalog \cite{sxs, 2013PhRvL.111x1104M}. 
We find that for BBH systems with misaligned spins, and that exhibit precession during their inspiral phase, a model consisting only of overtones (with or without mirror modes) \emph{cannot} be reliably applied from the peak of the 22 strain. A more conservative ringdown start time corresponding to the peak of the energy flux improves reliability, but we still see significant variation in performance across different simulations. The introduction of a higher harmonic (QNMs with $\ell > 2$) to the overtone model helps to reduce this variation, hinting at the importance of mode mixing.

Previous studies have also focused on using full NR simulations to test ringdown models. In this paper we briefly investigate the use of surrogates, which provide an opportunity to test models over a continuous parameter space. We find caution should be taken, particularly for surrogates of precessing systems, due to errors in the surrogate waveforms.

In section \ref{aligned-spin-section} we reproduce some important results from \cite{overtones, mirror_modes}, which are later compared with those for precessing systems in section \ref{misaligned-spin-section}. With precessing systems, it is necessary to perform a rotation to account for the fact that the spin of the remnant BH will not be aligned with initial coordinate axes used to set up the simulation; the procedure for doing this is also discussed in section \ref{misaligned-spin-section}. In section \ref{surrogate-section} we comment on the use of NR surrogates to test ringdown models. Finally, concluding remarks are presented in section \ref{sec:discussion}. Throughout, we use natural units in which $G=c=1$.

\section{Aligned-spin Systems}\label{aligned-spin-section}

As well as reproducing important results for spin-aligned systems \cite{overtones, mirror_modes}, this section introduces QNM modelling and describes our numerical fitting procedure. 

The GW signal far from a source of mass $M$ can be expanded in spin-weight $s=-2$ spherical harmonics as
\begin{equation}\label{YlmExpansion}
    h(t,r,\theta,\phi) = \frac{M}{r} \sum_{\ell = 2}^{\infty} \sum_{m = -\ell}^{\ell} h_{\ell m}(t) ~ {}_{-2}Y_{\ell m}(\theta, \phi).
\end{equation}
The $h_{\ell m}(t)$ coefficients are referred to as the spherical harmonic modes of the GW signal.
The ${\ell=|m|=2}$ modes are typically largest, while the remaining ``higher modes'' are generally subdominant.
The output of an NR simulation usually includes the first few modes (e.g.\ $\ell \leq 8$) with the asymptotic radial dependence scaled out.
The spherical harmonic modes are defined with respect to a particular frame at infinity.
This frame is chosen to be one in which the centre-of-mass of the system is at rest at some initial time; but this still leaves freedom to perform an overall rotation.
By convention, the \emph{NR frame} $(\theta,\phi)$ is uniquely fixed by requiring that initially the two component BHs are located on the $x$-axis and the orbital angular momentum, $\vb*{L}$, points along the $z$-axis.

At late times ($t \geq t_0$, where $t_0$ is to be determined) the signal is modelled as a sum of QNMs.
We note that as QNMs are not \emph{complete}, in the sense of being derivable from a self-adjoint operator, this model is necessarily an approximation.
The most general QNM ringdown model is a sum over $(\ell,m,n)$ including both the regular ($\omega_{\ell m n}$) and the mirror ($\omega'_{\ell m n}$) mode frequencies (see, e.g.\ \cite{bcw}),
\begin{widetext}
\begin{align}\label{general_ringdown}
    h(t,r,\theta',&\phi') =
    \frac{M_f}{r} \sum_{\ell=2}^{\infty}\sum_{m=-\ell}^{\ell}\sum_{n=0}^{\infty} \left[ C_{\ell m n} e^{-i \omega_{\ell m n}(t-t_0)} ~ {}_{-2}S_{\ell m n}(\theta',\phi') + C'_{\ell m n} e^{-i \omega'_{\ell m n}(t-t_0)} ~ {}_{-2}S'_{\ell m n}(\theta',\phi') \right] ,
\end{align}
\end{widetext}
for $t\geq t_0$.
Here, ${}_{-2}S_{\ell m n}(\theta',\phi')$ are the spheroidal harmonics of spin weight $-2$, which are the most natural angular basis for the radiation produced by a perturbed Kerr BH. 
The prime on the second spheroidal harmonic is due to the different QNM frequency associated with the mirror modes.
They are related to the regular mode spheroidal harmonics by ${}_{-2}S'_{\ell m n}(\theta',\phi') = {}_{-2}S^*_{\ell (-m) n}(\pi-\theta',\phi')$ \cite{mirror_modes}.
This model is constructed in the \emph{ringdown frame} $(\theta', \phi')$ in which the remnant BH is at rest with its spin vector pointing along the positive $z$-direction (such a frame is unique up to an unimportant $\phi'$ rotation about the $z$-axis).
For aligned-spin BBH systems, which do not precess during the inspiral, the ringdown and NR frames remain aligned with each other (at least up to an overall sign; some systems with strongly negative component spins can exhibit ``spin flip'' where the final spin points in the negative $z$-direction).
For misaligned-spin systems the remnant spin can point in essentially any direction and the NR and ringdown frames are misaligned.
The ringdown frame will also be moving with respect to the NR frame as a result of the recoil, or \emph{kick}, from the anisotropic emission of GWs near merger. The kick direction serves to single out a preferred $\phi'$ direction in the ringdown frame.
The effects of the kick are neglected here; it is assumed that the NR and ringdown frames are related by a rotation (see section \ref{misaligned-spin-section}).

Following Giesler \emph{et al.} \cite{overtones}, the spherical harmonic modes of the ringdown signal can be modelled by writing each as a sum of $N$ overtones:
\begin{equation}\label{GieslerRD}
    h_{\ell m}^N(t) = \sum_{n=0}^N C_{\ell m n} e^{-i\omega_{\ell m n}(t-t_0)}, \quad \textrm{for} \quad t \geq t_0.
\end{equation}
This \emph{overtone} model is a restriction of the sum in Eq.~\ref{general_ringdown}, where overlaps between different harmonic $\ell$ indices (mode mixing) \cite{2014PhRvD..90f4012B} as well as mirror modes are neglected. 
As in \cite{overtones}, we model each spherical harmonic mode individually as a sum of QNMs. An alternative approach would be to model several spherical harmonic modes simultaneously with a shared set of QNM amplitudes which might give improved fits, especially when mode mixing is significant (see, e.g.\ \cite{2020PhRvD.102b4027C}).
In \cite{overtones}, the efficacy of this model for $l=m=2$ was demonstrated by performing least squares fits to the $h_{22}(t)$ mode for a selection of aligned-spin SXS simulations. The authors note that this was also verified for other values of $(\ell,m)$.

The overtone model in Eq.~\ref{GieslerRD} contains $2(N+1)$ free parameters in the complex amplitudes, $C_{\ell m n}$, plus the two parameters $M_f,\; \chi_f$ that determine the $\omega_{\ell m n}$ frequencies.
All of these parameters depend on the properties of the progenitor binary, but we do not study these dependencies here.
We now briefly describe the numerical procedure used to search over these parameters and obtain a least-squares fit with the NR strain data $\vb{d}=h_{\ell m}(t)$.
Writing the difference between $\vb{d}$ and the overtone model in Eq.~\ref{GieslerRD} in its discretely sampled form (at times $t_0,\, t_1,\, \dots,\, t_{K-1}$) gives a linear matrix equation (we temporarily drop $\ell$ and $m$ indices for clarity),
\begin{widetext}
\begin{equation}\label{eq:linalg_fit}
\renewcommand*{\arraystretch}{1.5}
    ||\vb{d}-h^{N}_{\ell m}(t)|| = \norm{\mqty(d(t_0) \\ d(t_1) \\ \vdots \\ d(t_{K-1})) - 
    \mqty(e^{-i\omega_0(t_0-t_0)} & e^{-i\omega_1(t_0-t_0)} & \cdots & e^{-i\omega_N(t_0-t_0)} \\ 
          e^{-i\omega_0(t_1-t_0)} & e^{-i\omega_1(t_1-t_0)} & \cdots & e^{-i\omega_N(t_1-t_0)} \\ 
          \vdots & \vdots & \ddots & \vdots \\ 
          e^{-i\omega_0(t_{K-1}-t_0)} & e^{-i\omega_1(t_{K-1}-t_0)} & \cdots & e^{-i\omega_N(t_{K-1}-t_0)}) 
    \mqty(C_0 \\ C_1 \\ \vdots \\ C_N)}.
\end{equation}
\end{widetext}

Our fitting algorithm minimises the sum-of-the-squares of the fit residuals, which is the quantity in Eq.~\ref{eq:linalg_fit}. 
We find it convenient to treat the remnant property parameters $M_f$ and $\chi_f$ differently from the excitation amplitudes $C_n$. 
First, a discrete 2-dimensional numerical grid of values for 
$M_f$ and $\chi_f$ is constructed.
At each point on this grid, we consider varying only the complex amplitudes $C_n$. 
Eq.~\ref{eq:linalg_fit} turns this minimisation problem into a linear algebra problem that can be efficiently solved with, for example, \texttt{numpy.linalg.lstsq} \cite{numpy}.
Finally, the point of the grid with the lowest overall value for the sum-of-the-squares of the fit residuals is chosen.

\begin{figure}[t]
    \centering
    \includegraphics[width=\columnwidth]{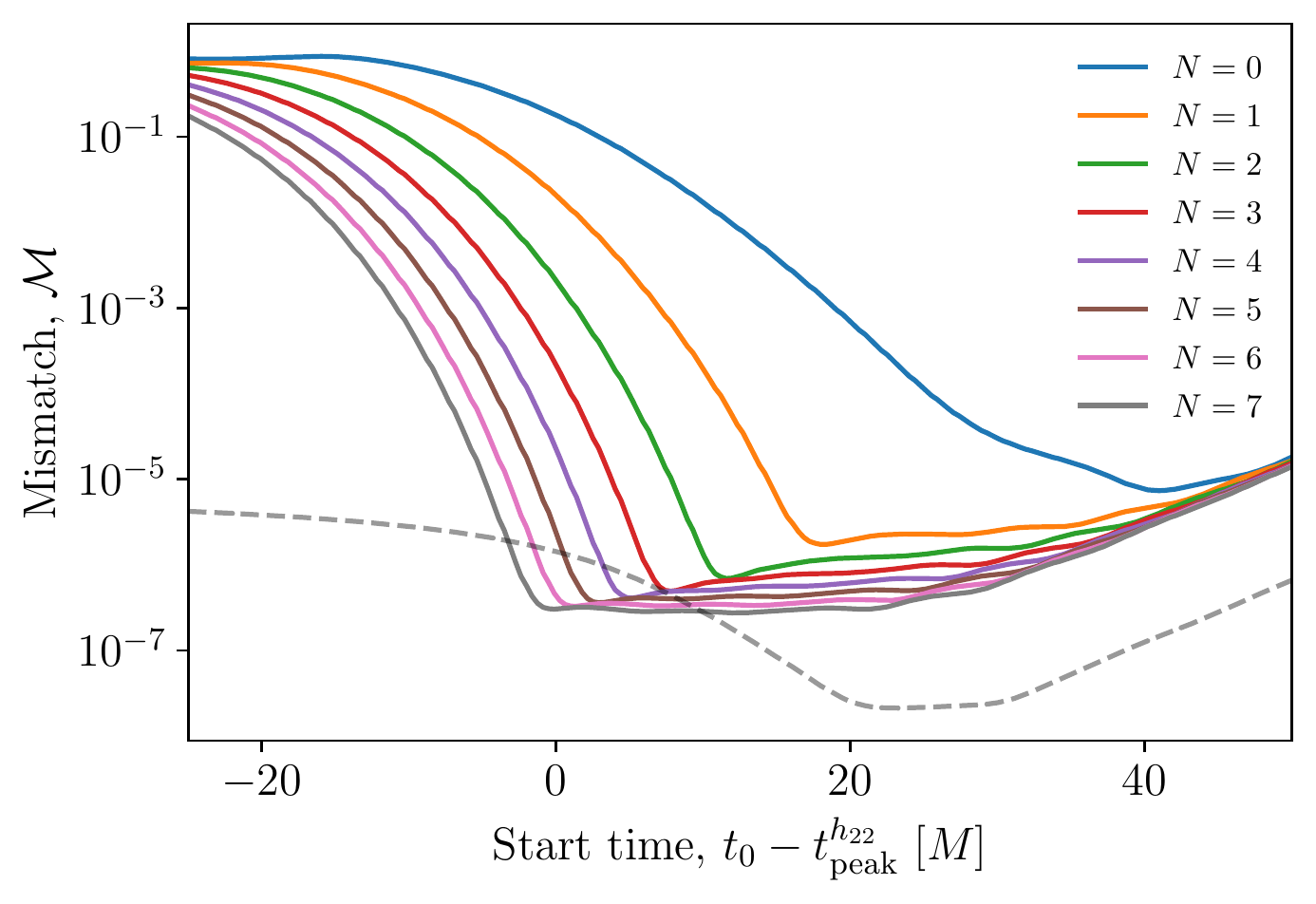}
    \caption{ \label{305_mismatch_vs_t0}
    The mismatch for the overtone model (Eq.~\ref{GieslerRD}) when fitting to the NR simulation SXS:BBH:0305 as a function of ringdown start time $t_0$. When using only a single QNM (the fundamental $(\ell,m,n)=(2,2,0)$) the start time that gives the lowest mismatch with the NR data is well after the merger (the rising mismatch at late times is a numerical artefact). However, reproducing the results from \cite{overtones}, we find that by including $N=7$ overtones the GW signal can be modelled using QNMs starting from as early as the peak strain. The dashed grey curve shows the estimate of the error in the underlying NR simulation and is described in appendix \ref{NR_error_appendix}.}
\end{figure}

\begin{figure}[t]
    \centering
    \includegraphics[width=\columnwidth]{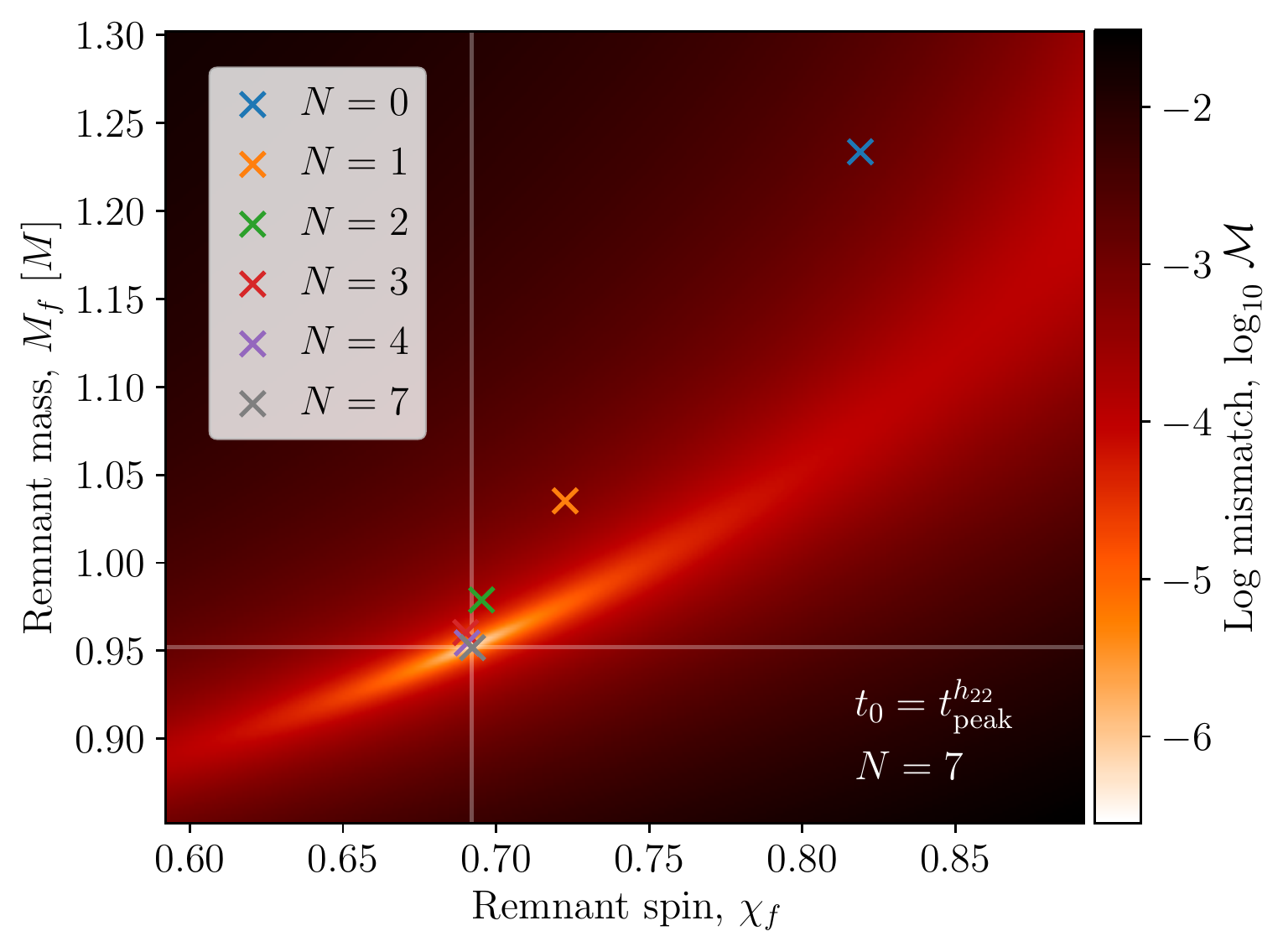}
    \caption{ \label{305_remnant_error}
    The recovery of the remnant properties for the overtone model (Eq.~\ref{GieslerRD}) when fitting to the NR simulation SXS:BBH:0305 starting from the peak in the $h_{22}$ strain.
    The heat map shows the mismatch for the fit with $N=7$ overtones, which shows a pronounced minimum close ($\epsilon=3.4\times 10^{-4}$) to the true remnant parameters (indicated by the horizontal and vertical lines).
    The sequence of crosses shows the locations of the minima for fits performed with the overtone model using different values of $N$, all using the same start time (the cross colours correspond to the colours used in Fig.~\ref{305_mismatch_vs_t0}; crosses for $N=5$ and 6 are omitted to avoid crowding the plot, but they converge towards the true remnant parameters). 
    If we choose a different ringdown start time for each $N$ corresponding to the mismatch minima in Fig.~\ref{305_mismatch_vs_t0}, we do see a reduction in $\epsilon$ for the lower $N$ models, however the $N=7$ model with $t_0=t_{\rm peak}^{h_{22}}$ remains the best performing model.
    } 
    \label{305_epsilon_grid}
\end{figure}

Once the least-squares fit to the data has been obtained, the quality, or \emph{goodness-of-fit}, is quantified via the mismatch and the error on the remnant parameters.
The \emph{mismatch} between signals $h_1$ and $h_2$ is defined as
\begin{equation}\label{mismatch}
    \mathcal{M} = 1 - \frac{\abs{ \braket{h_1}{h_2} }}{\sqrt{\braket{h_1}\braket{h_2}}},
\end{equation}
where we use the following complex inner product \cite{1999JMP....40..980N}
\begin{equation} \label{eq:inner_prod}
    \braket{h_1}{h_2} = \int_{t_0}^T h_1(t) h^*_2(t) ~ \dd t.
\end{equation}
We integrate from the ringdown start time, $t_0$, to an upper limit $T$ chosen such that the whole ringdown is captured (we use $T = t_0 + 100M$).
When fitting models with very small mismatches, the finite accuracy of the NR simulations must be considered; this is discussed in appendix \ref{NR_error_appendix}.
As noted in \cite{overtones}, a small mismatch is not sufficient by itself to justify the model.
The overtone model contains more parameters as $N$ is increased, and it is necessary to check for over-fitting.
To address this, we check to see if the remnant BH properties are correctly recovered by the model. 
The combined error on the remnant mass and spin is quantified by \cite{overtones}
\begin{equation} \label{eq:epsilon}
    \epsilon = \sqrt{ \qty( \frac{\delta M_f}{M} )^2 + \qty( \delta\chi_f )^2 },
\end{equation}
where $\delta M_f = M_{\mathrm{best fit}} - M_f$, and $\delta \chi_f = \chi_{\mathrm{best fit}} - \chi_f$. 
The best fit values are those which minimise the mismatch, while the true values are taken from the metadata for the SXS simulation.
A ringdown model can be said to perform well if it yields small values for \emph{both} $\mathcal{M}$ and $\epsilon$.

\begin{figure*}[t]
    \centering
    \includegraphics[width=2\columnwidth]{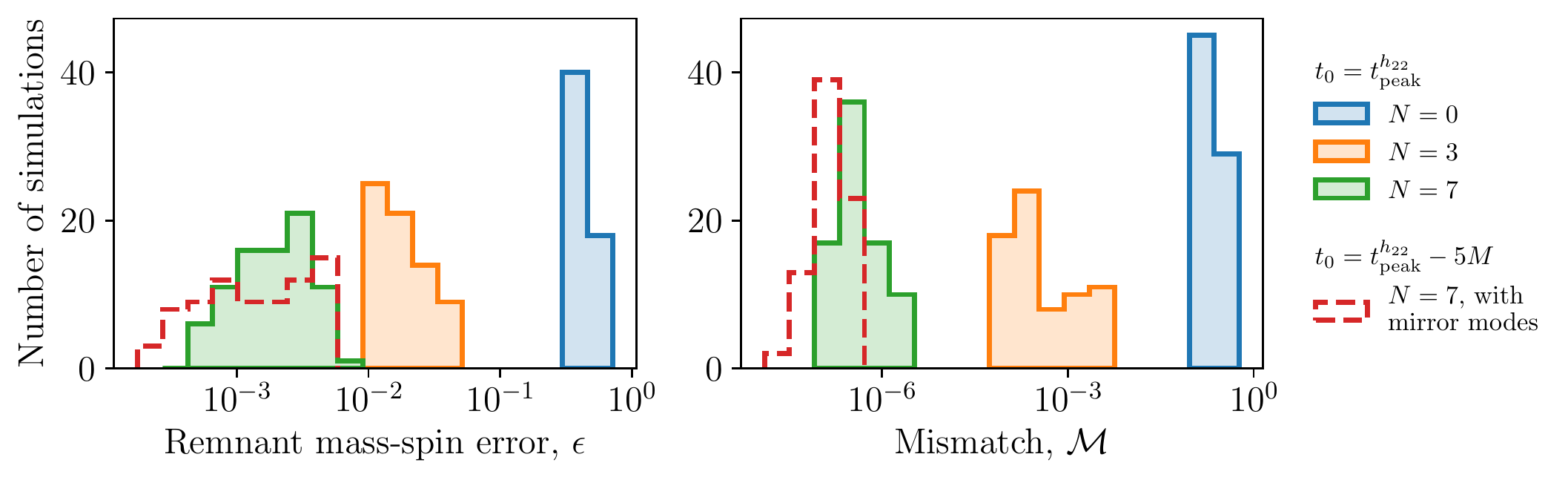}
    \caption{Left: histograms of the mass-spin remnant error $\epsilon$ from an overtone model fit to 85 aligned-spin SXS simulations for several different overtone numbers $N$. 
    Right: histograms of the mismatch from a fit with the true remnant mass and spin parameters, with the same overtone models and SXS simulations as in the left histogram.
    The solid histograms show results from fits performed starting at the peak of the $h_{22}$ mode with $N$ overtones of the fundamental $\ell = m = 2$ mode.
    The red dashed line shows results from a $N=7$ model that also includes mirror modes (see section \ref{subsec:mirror_modes}) and was fitted with a ringdown starting $5M$ before the peak in the strain.}
    \label{aligned_spin_epsilon_hist}
\end{figure*}

Following \cite{overtones}, we now apply these ideas to the simulation SXS:BBH:0305 \cite{2016CQGra..33x4002L, sxs_catalog}.
This simulation has source parameters consistent with GW150914 and was originally chosen to demonstrate the success of the overtone model. 
Fig.~\ref{305_mismatch_vs_t0} shows the mismatch values obtained with the overtone model when using the true values of $M_f$ and $\chi_f$.
With $N=7$ (that is, eight QNMs = the fundamental + seven overtones) the $h_{22}(t)$ mode can be fitted all the way back to the time of its peak amplitude, $t_{\mathrm{peak}}^{h_{22}}$, while still achieving the smallest possible mismatch.
Using a smaller number of overtones requires a later choice for the start time to achieve the smallest possible mismatch.
In addition to giving a small ($\sim 10^{-6}$) mismatch, the $N=7$ overtone model, with $t_0 = t_{\mathrm{peak}}^{h_{22}}$, also achieves this minimum mismatch with the correct values for the remnant properties; this is shown by the heat map in Fig.~\ref{305_epsilon_grid} where the values of $M_f$ and $\chi_f$ are now allowed to vary. 
We find, for the $N=7$ model, a remnant error $\epsilon = 3.4 \cross 10^{-4}$. 
Importantly, this is larger than the NR error on the remnant properties (which is estimated to be $\epsilon_{\mathrm{NR}} = 2.1 \cross 10^{-5}$, see appendix \ref{NR_error_appendix} for details). This confirms that this is really the true scale of the bias in the inferred remnant parameters when using the overtone model, and not just the numerical noise floor in the NR simulation.
Again, using a smaller number of overtones and starting the ringdown as early as $t_{\mathrm{peak}}^{h_{22}}$ gives inferior results with the minimum in the mismatch being biased away from the true parameters.
The results in Figs.~\ref{305_mismatch_vs_t0} and \ref{305_remnant_error} show that the overtone model performs well for SXS:BBH:0305 (i.e.\ yields small $\mathcal{M}$ and $\epsilon$) even when starting the ringdown as early as the peak in the strain.

In order to see how robust the conclusions drawn from SXS:BBH:0305 are in general, the calculations of $\epsilon$ and $\mathcal{M}$ were repeated for a wider selection of SXS simulations. Following \cite{overtones}, we consider only aligned-spin simulations with initial spin magnitudes $|\vb*{\chi}_{1,2}| = \chi_{1,2} < 0.8$, and mass ratios $q < 8$. We also require that the $z$-component of $\vb*{\chi}_f$ is greater than zero, which eliminates the ``spin flip'' systems. The simulations were chosen in the ID range SXS:BBH:1412 to SXS:BBH:1513, as these cover a range of initial spin magnitudes and mass ratios.
After applying these cuts, this left 85 spin-aligned SXS simulations in our test population.
For each simulation, fits were performed using the overtone model with $N=0$, 3, and 7 and with a start time of $t_0 = t_{\mathrm{peak}}^{h_{22}}$. 
The results are shown in Fig.~\ref{aligned_spin_epsilon_hist}. 
We see distributions similar to those in Fig.~3 of \cite{overtones}. 
The inclusion of additional overtones systematically shifts the entirety of both the $\epsilon$ and $\mathcal{M}$ histograms to smaller values.
We note, as it will become important later, that the worst cases in these histograms improve, along with the median values.
This demonstrates that, when using the overtone model on systems with aligned spins, the ringdown reliably starts as early as the peak in the $h_{22}(t)$ mode of the strain.

\subsection{Mirror Modes} \label{subsec:mirror_modes}

For a given $\ell$, $m$ and $n$, the equations governing QNM frequencies allow two solutions: one, $\omega_{\ell m n} = 2\pi f_{\ell m n} - i/\tau_{\ell m n}$, with a positive real part; and another, $\omega'_{\ell m n} = 2\pi f'_{\ell m n} - i/ \tau'_{\ell m n}$, with negative real part \cite{mirror_modes, bcw}.
The frequencies of the mirror modes $\omega'_{\ell m n}$ are related to the regular modes $\omega_{\ell m n}$ by
\begin{equation}\label{mirrorsymmetry}
    f'_{\ell m n} = -f_{\ell -m n}, \quad \tau'_{\ell m n} = \tau_{\ell -m n} \nonumber
\end{equation}
\begin{equation} 
    \quad\Rightarrow\quad \omega'_{\ell mn} = - \omega_{\ell -mn}^*.
    \label{eq:sym_mirror_modes_conj}
\end{equation}

A new ringdown model which explicitly includes the mirror modes can be written as
\begin{align}
    h_{\ell m}^{N,\, {\rm mirror}}(t) = \sum_{n=0}^N \Big[& C_{\ell m n} e^{-i \omega_{\ell m n}(t-t_0)} \\ &+ C'_{\ell m n} e^{-i \omega'_{\ell m n}(t-t_0)} \Big]\quad \textrm{for} \quad t \geq t_0. \nonumber
\end{align}
This \emph{mirror mode} model is an extension of the overtone model in Eq.~\ref{GieslerRD};
if $C'_{\ell m n} = 0$ the mirror modes aren't excited and we recover the previous overtone model. 
This model has twice as many free parameters as the overtone model; $4(N+1)$ in the complex amplitudes, plus the two remnant parameters $M_f,\; \chi_f$.
The mirror mode model is still a restriction of the full sum in Eq.~\ref{general_ringdown} as overlaps between modes with different $\ell$ indices (i.e.\ mode mixing) are still not included.
Substituting for $\omega'_{\ell m n}$ using the conjugate symmetry property in Eqs.~\ref{eq:sym_mirror_modes_conj}, we can rewrite the mirror mode model in the form
\begin{align} \label{mirror_ringdown}
   h_{\ell m}^{N,\, {\rm mirror}}(t) = \sum_{n=0}^N \Big[& C_{\ell m n} e^{-i \omega_{\ell m n}(t-t_0)} \\ &+ C'_{\ell m n} e^{i \omega^*_{\ell -m n}(t-t_0)} \Big]\quad \textrm{for} \quad t \geq t_0. \nonumber
\end{align}
It is this form of the mirror mode model that was implemented.

As was shown in \cite{mirror_modes}, the inclusion of mirror modes can improve the ringdown modelling of aligned-spin systems. In particular, the ringdown can be considered to start even earlier in the waveform, whilst still recovering the correct remnant properties. We confirm this here by repeating the above analysis for the same set of spin-aligned SXS simulation, but now using the mirror mode model in Eq.~\ref{mirror_ringdown} with $N=7$ and an earlier choice for the ringdown start time, $t_0 = t_{\mathrm{peak}}^{h_{22}} - 5M$.
Although \cite{mirror_modes} demonstrated the mirror mode model starting $10M$ before the peak in the $h_{22}$ strain, we adopt a more conservative choice of $5M$.
The results are shown in Fig.~\ref{aligned_spin_epsilon_hist} plotted using a dashed line. 
The addition of mirror modes gives a small improvement in the mismatch, but this is to be expected with the increased number of parameters.
However, the $\epsilon$ histogram shows that the overall performance of the mirror mode model is comparable to that of the $N=7$ overtone model, despite the use of an earlier start time.

\section{Misaligned-spin Systems}\label{misaligned-spin-section}

The analyses in section \ref{aligned-spin-section}, and in the previous studies \cite{1998PhRvD..57.4535F, 2007PhRvD..76f4034B, 2012PhRvD..85b4018K, 2014PhRvD..90l4032L, 2018PhRvD..98j4020C, 2018PhRvD..97j4065B, overtones, 2020PhRvD.101j4005O, mirror_modes}, was limited to BBH systems with component spins that are aligned with the orbital angular momentum, $\vb*{L}$.
This is a potentially serious limitation as misaligned spins are expected to be a generic feature of astrophysical BBHs.
Misaligned spins generally lead to precession of the orbital plane during the inspiral phase of the evolution and a richer phenomenology in the GW signals \cite{1994PhRvD..49.6274A}.
Several of the GW detections already show signs of precession, both individually \cite{2020arXiv201014527A, 2020arXiv201111948G} and when considered as a population \cite{2020arXiv201014533T}.
Most notably for our present purposes, the very high-mass system GW190521 \cite{2020PhRvL.125j1102A, 2020ApJ...900L..13A} might show some signs of precession while also having a high fraction of the observable signal-to-noise ratio in the ringdown. 
In this section we investigate the effect of precession on the modelling of the ringdown by repeating analyses like those in section \ref{aligned-spin-section}, but now on precessing NR simulations.

The general ringdown signal in Eq.~\ref{general_ringdown} is written in the ringdown frame $(\theta', \phi')$ which is aligned with the remnant spin vector (i.e.\ $\vb*{\chi}_f$ points along the positive $z$-direction in this frame). For BBH systems with misaligned spins that undergo precession, the ringdown frame differs from the frame typically used in NR simulations $(\theta, \phi)$ which is aligned with $\vb*{L}$ at some arbitrary start time in the inspiral.
These two frames are related by a rotation, $\mathbf{R}$.

The direction from which a GW source is viewed affects the observed signal 
(e.g.\ you see circularly/linearly polarised GWs with a larger/smaller amplitude when viewing parallel/perpendicular to $\vb*{L}$).
These differences in the GW signals also manifest themselves at the level of individual modes as amplitude modulations. The frame in which the expansion is performed affects the values of the spherical harmonic modes. 
The GW signal, originally given in the NR frame in Eq.~\ref{YlmExpansion}, can be re-expanded in the ringdown frame as follows:
\begin{equation}\label{hprimedecomp}
    h'(t, r, \theta', \phi') = \frac{M}{r} \sum_{\ell = 2}^{\infty} \sum_{m = -\ell}^{\ell}  h'_{\ell m}(t) ~ {}_{-2}Y_{\ell m}(\theta', \phi').
\end{equation}
When analysing the ringdown, it is most natural to use the spherical harmonic modes in the ringdown frame, $h'_{\ell m}(t)$, as these are adapted to the remnant BH. 
In particular, we will focus on modelling the $\ell = m = 2$ spherical harmonic mode in the ringdown frame, $h'_{22}(t)$.

\begin{figure}[b]
    \centering
    \includegraphics[width=\columnwidth]{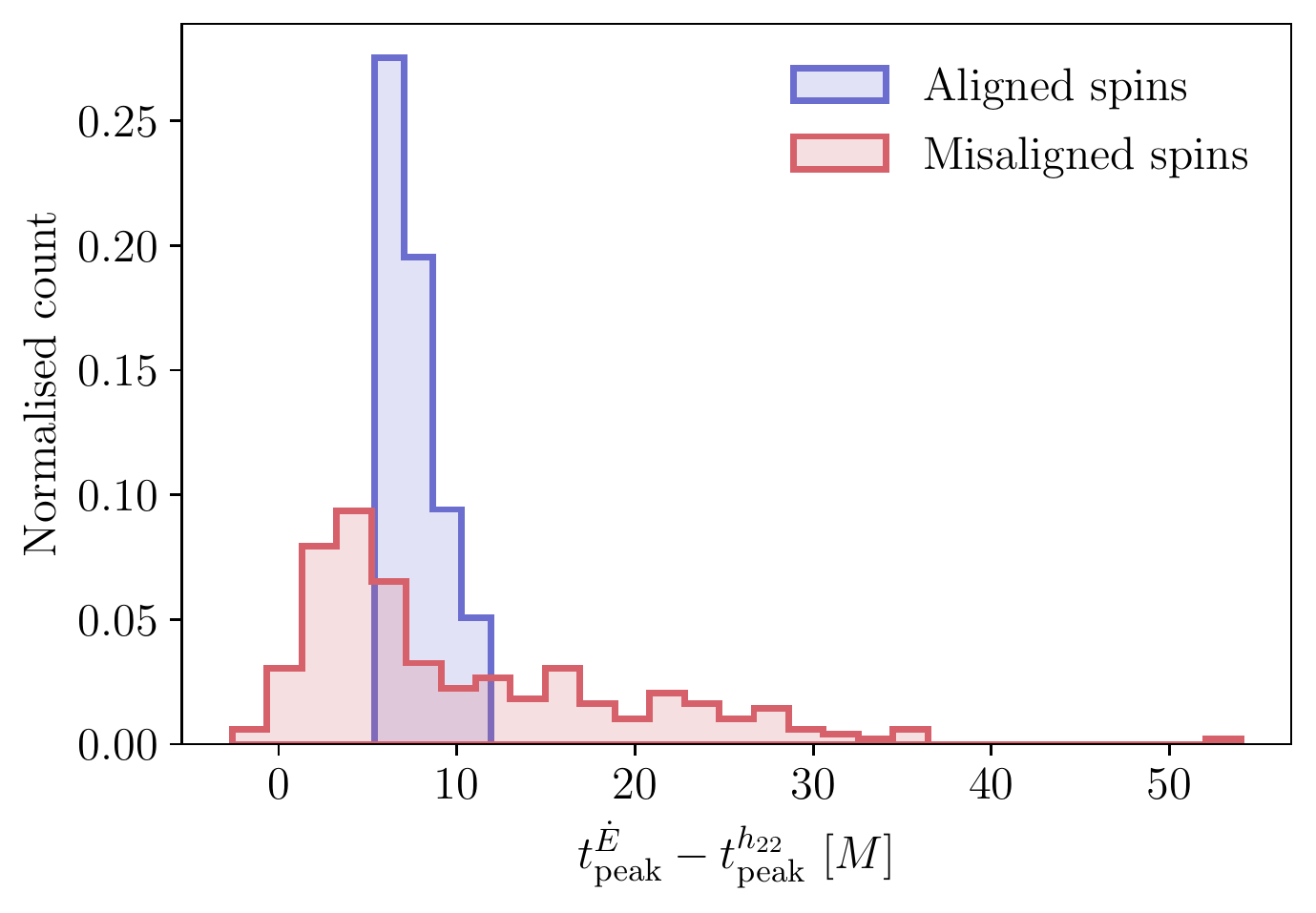}
    \caption{ \label{tEdot-t22} 
    Histogram of the differences between the two possible start times considered in section \ref{misaligned-spin-section}: the peak of the (rotated) strain mode $h'_{22}$, and the peak of the GW energy flux.
    The normalised distribution of the differences between these times is shown both for the 85 spin-aligned systems used in section \ref{aligned-spin-section} and for the 252 precessing simulations considered in section \ref{misaligned-spin-section}. 
    The peak of the flux almost always occurs later than the peak of the strain, making this a more conservative choice for the ringdown start time. 
    We note that there is a much greater variation amongst the population with misaligned spins.
    }
\end{figure}

For aligned-spin systems, the $h_{2\pm2}$ are usually the dominant modes in the sum in Eq.~\ref{YlmExpansion}. This is related to the fact that the GW signal amplitude is largest when viewed along the direction of the orbital angular momentum: $\vb*{L}$ or $-\vb*{L}$. For misaligned-spin systems undergoing precession, other modes become important. This in turn is related to the constantly changing direction of the orbital angular momentum, $\vb*{L}(t)$.
Changing into the non-inertial, \emph{coprecessing} frame in which $\vb*{L}$ always points along the $z$-direction has been found to account for most precessional effects and makes the precessing waveform remarkably similar to a non-precessing one.
This transformation into the coprecessing frame has been successfully used to help model the full inspiral-merger-ringdown waveforms for precessing systems \cite{2011PhRvD..84b4046S,2012PhRvD..86j4063S} in the context of phenomenological \cite{2014PhRvL.113o1101H, 2019PhRvD.100b4059K, 2020arXiv200406503P}, effective-one-body \cite{2014PhRvD..89h4006P, 2020PhRvD.102d4055O} and NR surrogate \cite{2017PhRvD..95j4023B, 2017PhRvD..96b4058B, 2019PhRvR...1c3015V} modelling.
There is an analogy with the approach taken here for the modelling of the ringdown. In order to simplify the task, we choose to work in a frame adapted to final spin angular momentum of the remnant, $\vb*{\chi}_f$.
Although, in our case, the rotation required to get into this frame is not time dependent and our chosen frame is therefore inertial.

\begin{figure}[t]
    \centering
    \includegraphics[width=\columnwidth]{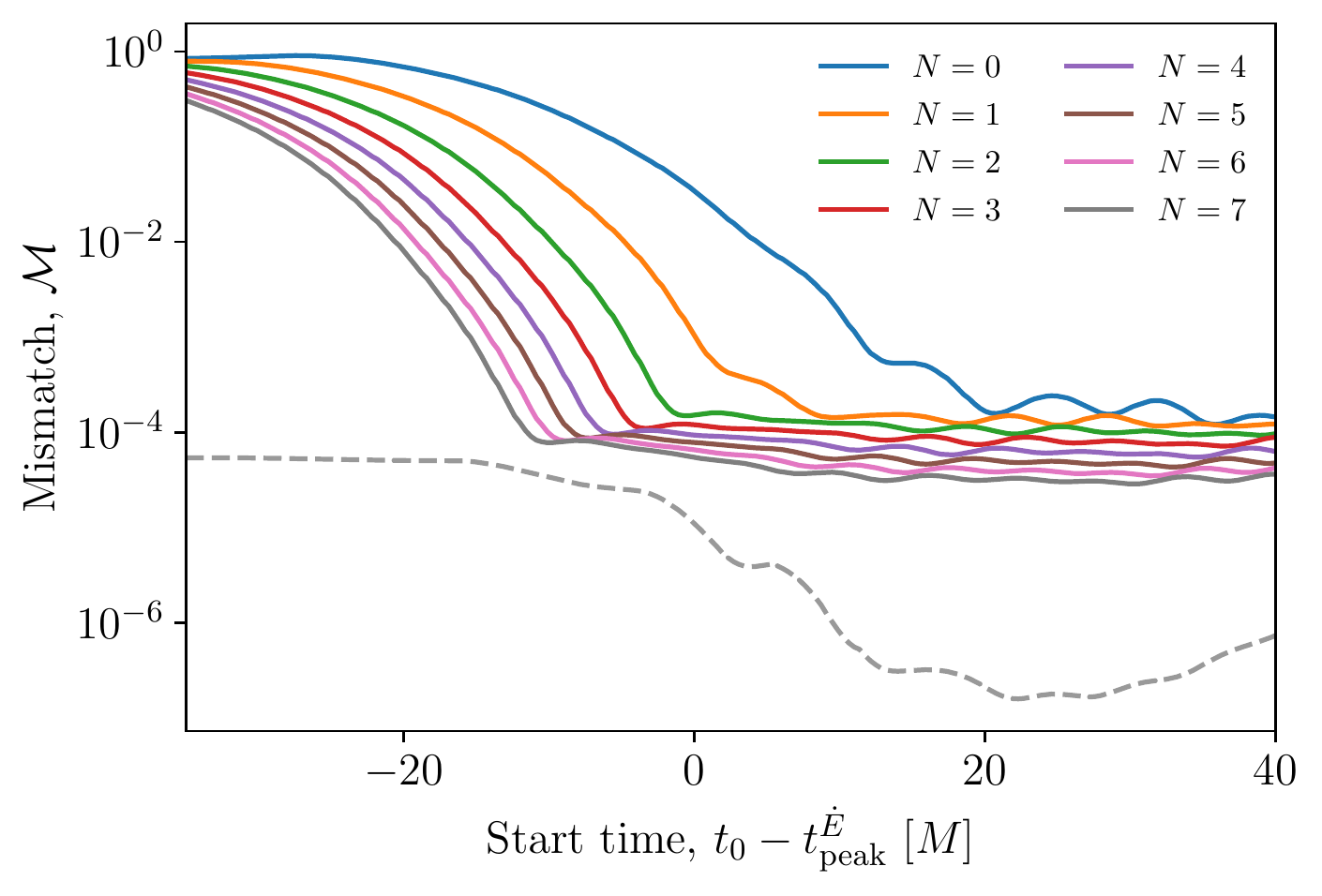}
    \caption{ \label{1856_mismatch_vs_t0}
    The mismatch for the overtone model (Eq.~\ref{GieslerRD}) when fitting to the NR simulation SXS:BBH:1856 as a function of ringdown start time $t_0$. When using $N=7$ overtones, the lowest mismatch is achieved starting slightly ($\sim 10M$) before the peak in the GW energy flux.
    However, the minimum mismatch is $\sim 100$ times larger than that obtained for the example spin-aligned system SXS:BBH:0305 in Fig.~\ref{305_mismatch_vs_t0}. The dashed grey curve shows the estimate of the error in the underlying NR simulation and is described in appendix \ref{NR_error_appendix}.
    }
\end{figure}

The spin-weighted spherical harmonics, ${}_{-2}Y_{\ell m}$, transform in a particularly simple manner under rotations.
Rotations have the effect of mixing together modes with different $m$ indices, but preserving the same $\ell$. 
The mixing coefficients in these transformations are the Wigner $D$-matrices $D^{\ell}_{\mu m} (\mathbf{R})$.
The transformation properties of the ${}_{-2}Y_{\ell m}$ functions under rotations means that the rotated $h'_{\ell m}$ modes in the ringdown frame are related to the $h_{\ell m}$ modes in the NR frame (included in the NR output, and as used in section \ref{aligned-spin-section}) by the sum (see, for example \cite{rotation, 2011PhRvD..84b4046S, 2011PhRvD..84l4002O})
\begin{align}\label{Yrotation_wignerD}
    h'_{\ell m}(t) &= \sum_{\mu = -\ell}^{\ell} D^{\ell}_{\mu m} (\mathbf{R}) ~ h_{\ell \mu}(t).
\end{align}
The rotation $\mathbf{R}$ can be obtained from the direction of the remnant BH spin vector (which is provided as metadata for all SXS simulations). Specifically, $\mathbf{R}$ is any rotation that maps the $z$-axis onto the final spin vector.

\begin{figure}[t]
    \centering
    \includegraphics[width=\columnwidth]{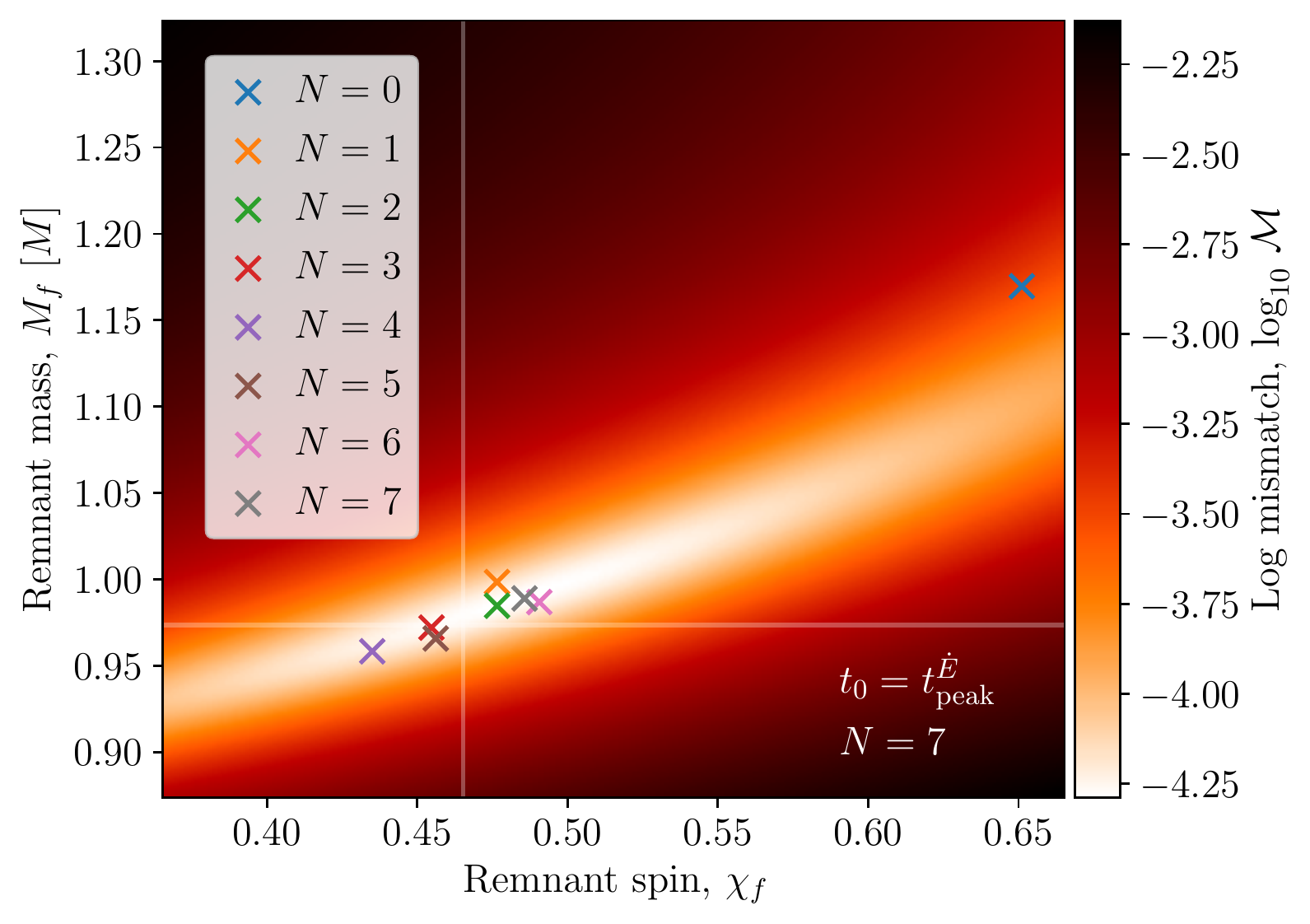}
    \caption{ \label{1856_epsilon_grid}
    The recovery of the remnant properties for the overtone model (Eq.~\ref{GieslerRD}) when fitting to the NR simulation SXS:BBH:1856 starting from the peak in the flux.
    The heat map shows the mismatch for the fit with $N=7$, while the crosses show the locations of the minima in the mismatch for fits performed with different values of $N$.
    The mismatch shows a much broader and less deep minimum than that seen for the spin-aligned system SXS:BBH:0305 in Fig.~\ref{305_epsilon_grid}.
    The minimum in the mismatch is also biased away from the true remnant parameters with $\epsilon=0.025$ for the $N=7$ fit.
    The sequence of crosses for fits with different values of $N$ also do not show the same convergent trend towards the true remnant parameters that was observed for SXS:BBH:0305 in Fig.~\ref{305_epsilon_grid}.
    }
\end{figure}

We now apply the overtone model to the ringdown of an example precessing simulation SXS:BBH:1856 \cite{2019PhRvR...1c3015V}. 
This simulation initially (at the reference time) has a mass ratio of $q=2.78$ and dimensionless spins $\vb*{\chi}_1=(0.18, -0.54, -0.45)$ and $\vb*{\chi}_2=(-0.12, -0.31, -0.031)$ on the heavier and lighter components respectively. This simulation was chosen because it exhibits strong precession effects visible as amplitude modulations in $h_{22}(t)$. The final spin vector is $\vb*{\chi}_f=(-0.03,-0.19,0.42)$ and the rotated mode $h'_{22}(t)$ was computed using Eq.~\ref{Yrotation_wignerD}.

\begin{figure*}[t]
    \centering
    \includegraphics[width=2\columnwidth]{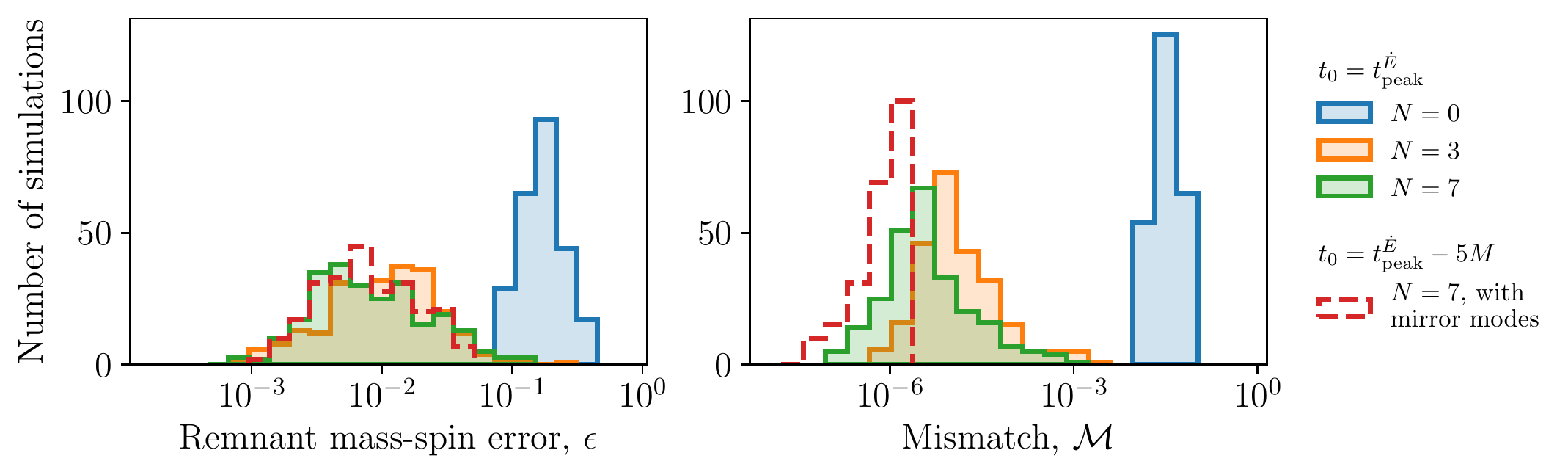}
    \caption{Left: histograms of the mass-spin remnant error $\epsilon$ from an overtone model fit to the rotated $h'_{22}$ modes of 252 misaligned-spin SXS simulations for several different overtone numbers $N$. 
    Right: histograms of the mismatch from a fit with the true remnant mass and spin parameters, with the same overtone models and SXS simulations as in the left histogram.
    The solid histograms show results from fits performed starting at the peak of the energy flux with $N$ overtones of the fundamental $\ell = m = 2$ mode.
    The red dashed line shows results from a $N=7$ model that also includes mirror modes and was fitted with a ringdown starting $5M$ before the peak in the energy flux.
    These histograms should be compared with those in Fig.~\ref{aligned_spin_epsilon_hist}; we note that the effect of precession is to (i) significantly broaden the histograms (i.e.\ the quality of the fit is much more varied) and (ii) to significantly degrade the quality of the fit for some systems.}
    \label{misaligned_spin_epsilon_hist}
\end{figure*}

The overtone model in Eq.~\ref{GieslerRD} was fitted to the rotated $\ell=m=2$ mode of the strain, $h'_{22}(t)$, in the same way as was done for the aligned-spin systems in section \ref{aligned-spin-section}.
There is some ambiguity in how to choose the ringdown start time $t_0$ in a way that gives as fair a comparison as possible with the non-precessing case.
We cannot use the peak of the $h_{22}(t)$ strain, as was done in section \ref{aligned-spin-section}, as this mode suffers from precession induced amplitude modulations. 
One option would be to use instead the peak of the rotated strain mode $h'_{22}(t)$.
However, we find that using the peak of the GW energy flux, $\dot{E}$, (which can be computed from the modes in either frame, see Eq.~3.8 in \cite{2008GReGr..40.1705R}) gives more consistent results between simulations. For example, some precessing configurations show a peak in the (rotated) strain relatively early in the signal, leading to poorer fits.
The use of the peak in the flux is also a conservative choice in the sense that $t_{\mathrm{peak}}^{\dot{E}} > t_{\mathrm{peak}}^{h_{22}}$ in almost all cases (see Fig.~\ref{tEdot-t22}).

Fig.~\ref{1856_mismatch_vs_t0} shows how the mismatch varies for SXS:BBH:1856 as a function of ringdown start time, for different values of $N$ in the overtone model Eq.~\ref{GieslerRD}.
With each additional overtone, the minimum mismatch is reached at an earlier time (the same behaviour as was seen in Fig.~\ref{305_mismatch_vs_t0}).
However, the values of the minimum mismatch are a factor of $\sim 100$ larger than those obtained in the aligned-spin case. 

The $N=7$ model achieves a minimum mismatch $\sim 10M$ before the time of peak GW energy flux. This is fairly typical behaviour among the misaligned-spin SXS simulations considered.
However, we note there is a much greater variety of possible behaviours for misaligned-spin systems than for the aligned-spin population. 
In order to emphasise this greater variation, in appendix \ref{appendix_a} we repeat the analysis in Figs.~\ref{1856_mismatch_vs_t0} and \ref{1856_epsilon_grid} for three more misaligned-spin simulations and highlight some of the observed differences. 
The greater variation amongst the misaligned-spin population has already been hinted at in Fig.~\ref{tEdot-t22}, where the spread of start times is greater than in the aligned-spin cases.

The heat map of Fig.~\ref{1856_epsilon_grid} shows the mismatch as a function of the remnant BH properties, for the $N=7$ model.
The coloured crosses indicate the mismatch minimum for different values of $N$. 
Comparing with Fig.~\ref{305_epsilon_grid}, we see the mismatch minimum is less pronounced than the aligned-spin case, which is probably contributing to the larger value of $\epsilon$ (for $N=7$ we find $\epsilon = 0.025$, which is much larger than the estimated numerical error $\epsilon_{\mathrm{NR}} = 8.6 \cross 10^{-5}$). 
In addition, the convergent behaviour with increasing $N$ is not present. For $N \geq 1$, all mismatch minima appear randomly distributed around the true remnant properties.
If we reproduce this figure with a earlier start time of $t_0 = t_{\mathrm{peak}}^{\dot{E}} - 10M$ (motivated by the time of minimum mismatch for $N=7$ in Fig.~\ref{1856_mismatch_vs_t0}), the heat map remains unchanged, and the value of $\epsilon$ recovered for $N=7$ is not significantly improved ($\epsilon = 0.013$). The earlier start time does cause the value of $\epsilon$ for $N \leq 3$ to increase significantly, which may be expected as we are now using a start time before those models reach a mismatch minimum. 

Following section \ref{aligned-spin-section}, we now extend this analysis to a wider selection of SXS simulations to investigate the robustness (or lack thereof) of this behaviour. We consider only misaligned-spin simulations, chosen such that the angle between the initial spins, $\chi_{\theta}$, satisfies $\pi/16 < \chi_{\theta} < 15\pi/16$. We again require initial spin magnitudes $\chi_{1,2} < 0.8$ and mass ratios $q<8$. The 252 simulations were chosen in the ID range SXS:BBH:1643 to SXS:BBH:1899, as these cover a range of mass ratios and initial spin configurations.

The results are shown in Fig.~\ref{misaligned_spin_epsilon_hist}. 
When compared to the $N=0$ model, the addition of three overtones reduces the remnant error and mismatch. 
However, the inclusion of additional overtones does not change the $\epsilon$ histogram, and produces only a minor reduction in the mismatch.
Comparing the $N=7$ histogram for $\epsilon$ to that found in Fig.~\ref{aligned_spin_epsilon_hist}, we see that, on average, $\epsilon$ increases by a factor of $\sim 10$ and, in the worst cases, by a factor of $\sim 20$ (however, the overtone model does still perform similarly well for a small fraction of simulations). 
The histograms for $\epsilon$ reflect the behaviour of Fig.~\ref{1856_epsilon_grid}, where models with $N \geq 1$ don't show systematic improvements. 
It would be interesting to investigate whether the binary parameters correlate with $\epsilon$, and if certain binary configurations are responsible for the largest remnant errors. We have performed preliminary studies which reveal no clear correlations of $\epsilon$ with either the amount of precession (quantified via $\chi_p$ \cite{2015PhRvD..91b4043S}) or the recoil velocity. We defer a more detailed study of this question to future work.

It was checked if using an earlier start time of $t_{\mathrm{peak}}^{\dot{E}} - 10M$ changed the recovered distribution on $\epsilon$. This choice was motivated by the location of the mismatch minimum typically seen for misaligned-spin simulations (e.g.\ see Fig.~\ref{1856_mismatch_vs_t0}). The results are shown in appendix \ref{misaligned_spin_fits_appendix}. It was found the $N=7$ model results did not significantly change. However, the $N=3$ and $N=0$ models performed worse.
Finally, we also note that all of the histograms are wider than those in Fig.~\ref{aligned_spin_epsilon_hist}. 
This may be due to mirror modes and/or higher harmonics having a more important role for precessing systems (see below). 

\begin{figure}[t]
    \centering
    \includegraphics[width=\columnwidth]{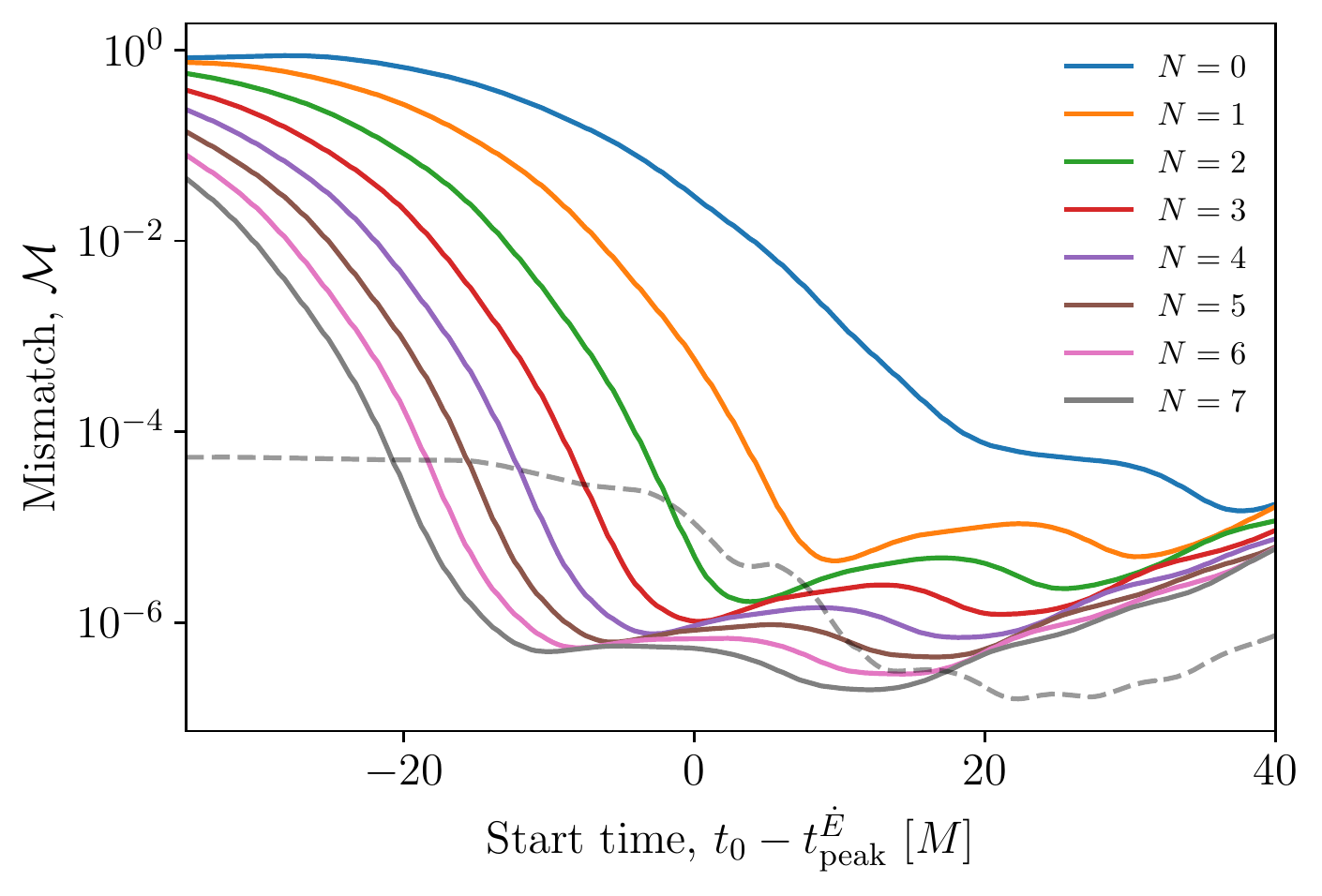}
    \caption{ \label{1856_mirror_mode_mismatch_vs_t0}
    The mismatch for the mirror mode model (Eq.~\ref{mirror_ringdown}) fitted to the NR simulation SXS:BBH:1856 as a function of ringdown start time $t_0$. Comparing with Fig.~\ref{1856_mismatch_vs_t0}, the locations of the mismatch minima are roughly unchanged in time, but the inclusion of mirror modes reduces the mismatch to values similar to those in Fig.~\ref{305_mismatch_vs_t0}. The dashed grey curve shows the estimate of the error in the underlying NR simulation and is described in appendix \ref{NR_error_appendix}.
    }
\end{figure}

\subsection{Mirror Modes} \label{subsec:misaligned_mirror_modes}

We repeat the population analysis with the $N=7$ mirror mode model, again shifting the ringdown start time back by $5M$ to make a clear comparison to Fig.~\ref{aligned_spin_epsilon_hist}. 
The results are shown by the red dashed lines in Fig.~\ref{misaligned_spin_epsilon_hist}. The histogram for $\epsilon$ doesn't reach values as high as the overtone model (with worst-case values of $\epsilon \sim 0.04$ compared to the overtone model's $\sim 0.2$), but otherwise has a broadly similar distribution.
However, there is a significant improvement on the recovered mismatch values. This is expected because of the large number of parameters. And, as discussed, this alone isn't enough to say the model is successful.

\begin{figure}[t]
    \centering
    \includegraphics[width=\columnwidth]{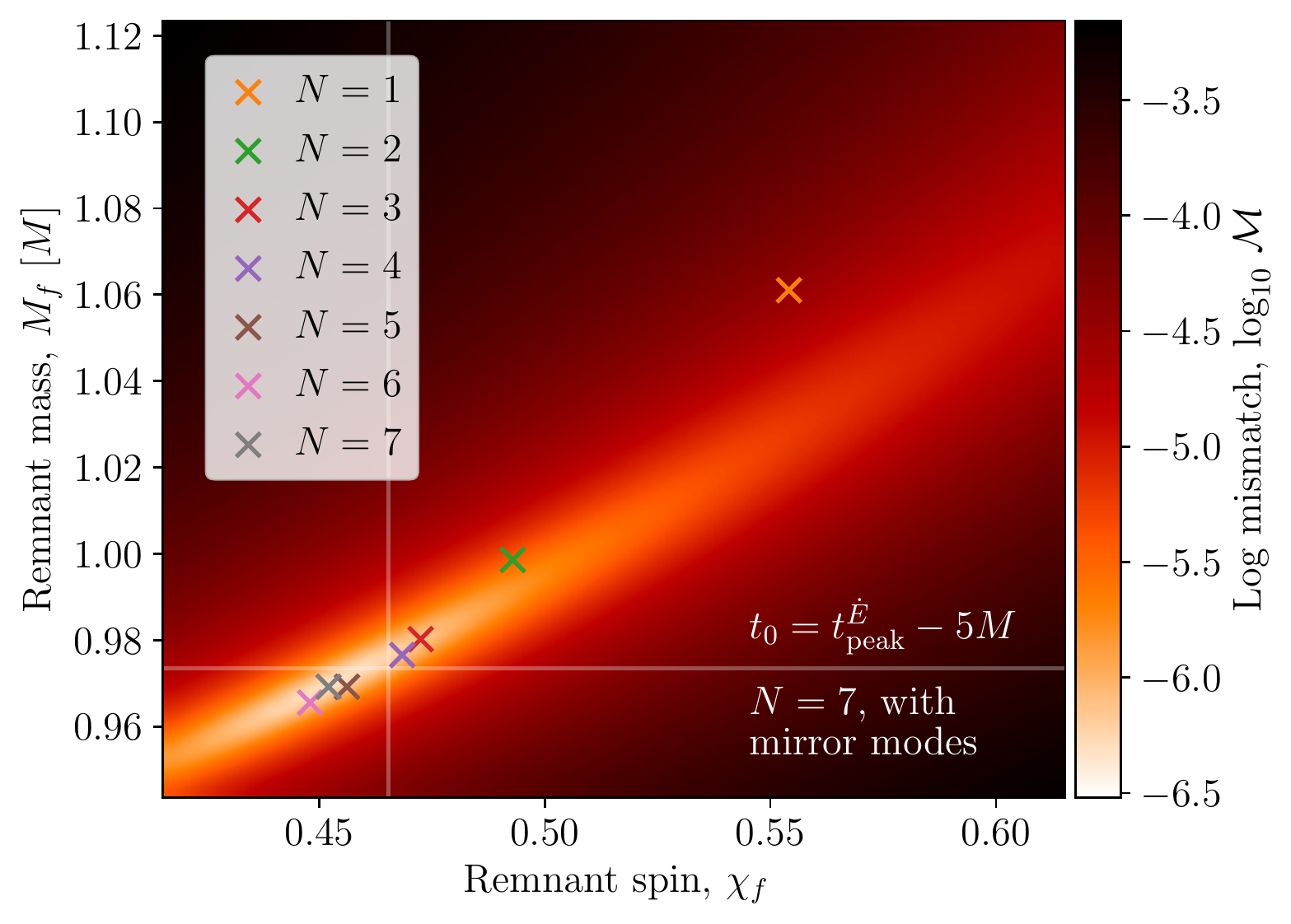}
    \caption{ \label{1856_mirror_mode_epsilon_grid}
    The recovery of the remnant properties for the mirror mode model (Eq.~\ref{mirror_ringdown}) when fitting to the NR simulation SXS:BBH:1856, starting from $5M$ before the peak of the flux. The heat map shows the mismatch for the fit with $N=7$, while the crosses show the locations of the minima in the mismatch for fits performed with different values of $N$ ($N=0$ lies outside the figure, and is not included for clarity). Comparing with Fig.~\ref{1856_epsilon_grid}, the inclusion of mirror modes sharpens the mismatch peak and achieves smaller mismatch values. However, when averaged across the population of precessing simulations, the mirror mode model doesn't give smaller values for the remnant error (see dashed curve in Fig.~\ref{misaligned_spin_epsilon_hist}). Here, $\epsilon = 0.014$ for the $N=7$ model.
    }
\end{figure}

Inspecting individual simulations, we see that the inclusion of mirror modes can make the mismatch minima in the mass-spin plane more pronounced (advantageous, as it reduces uncertainty on $\epsilon$). For example, Figs.~\ref{1856_mirror_mode_mismatch_vs_t0} and \ref{1856_mirror_mode_epsilon_grid} show how mirror mode fits perform for SXS:BBH:1856. 
We see significantly smaller mismatches, and a stronger mismatch peak around the true remnant properties. However, on average this does not translate to smaller values of $\epsilon$ for the $N=7$ model (as can be seen from the red dashed histogram in Fig.~\ref{misaligned_spin_epsilon_hist}). For SXS:BBH:1856, the $N=7$ model gives $\epsilon = 0.014$, which is not a significant improvement.

\begin{figure*}[t]
    \centering
    \includegraphics[width=2\columnwidth]{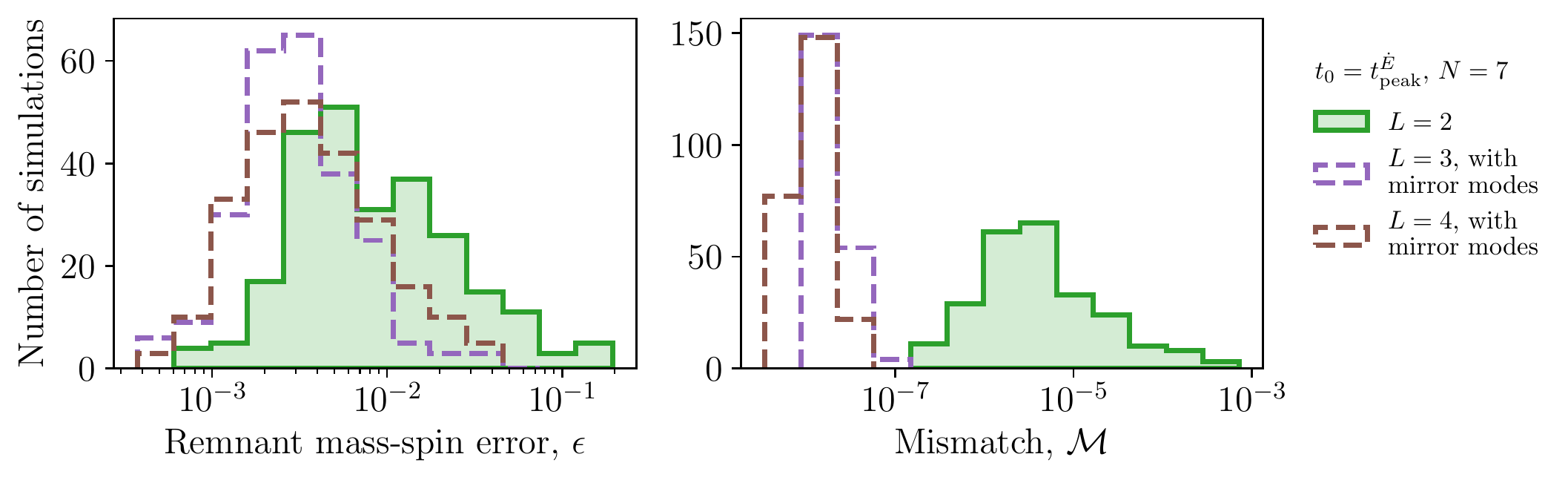}
    \caption{Left: histograms of the mass-spin remnant error $\epsilon$ from harmonic model fits (Eq.~\ref{full_ringdown}) to same 252 misaligned-spin SXS simulations used in Fig.\ref{misaligned_spin_epsilon_hist}. Shown (in dashed lines) are the $N=7$, $L=3$ and $L=4$ models with mirror modes. Also shown in green is the overtone model with $N=7$ and $L=2$ (no mirror modes); this is the same as the green histogram in Fig.~\ref{misaligned_spin_epsilon_hist} and is included here to aid comparison. Right: histograms of the mismatch from a fit with the true remnant mass and spin parameters, with the same models and SXS simulations as in the left histogram. The harmonic model, which includes many free parameters, achieves small mismatches but without significant improvement in the remnant error. We note that the inclusion of $L=4$ does not bring any additional improvements over $L=3$.}
    \label{misaligned_spin_epsilon_hist_harmonics}
\end{figure*}

To investigate whether the choice of ringdown start time could be contributing to the wider histograms seen in Figs.~\ref{misaligned_spin_epsilon_hist} and \ref{misaligned_spin_epsilon_hist_m10}, the behaviour of the mismatch heat maps (e.g.\ Figs.~\ref{305_epsilon_grid}, \ref{1856_epsilon_grid}, \ref{1856_mirror_mode_epsilon_grid}) with varying start time was explored for selected SXS simulations. 
Animations of ringdown fits with varying start time can be found at \cite{finch_eliot_2021_4538194}.
For the aligned-spin simulation SXS:BBH:0305, we see that the location of the mismatch minimum in the mass-spin plane settles on the true remnant properties for a sufficiently late choice of the start time ($t_0 \geq t_{\mathrm{peak}}^{h_{22}}$ for the $N=7$ overtone model). 
In addition, the mismatch minimum stays centred on the true remnant properties until numerical noise takes over.
For earlier choices of the start time, the $N=7$ overtone model gives biased values for the final mass and spin, see \cite{finch_eliot_2021_4538194}.
Applying the $N=7$ overtone model to the misaligned-spin simulation SXS:BBH:1856, we see that the location of the mismatch minimum moves around the mass-spin plane as start time is varied. Even at late times, it never settles on the true remnant properties.
The inclusion of mirror modes, as seen in Fig.~\ref{1856_mirror_mode_epsilon_grid}, narrows the mismatch minimum. The movement of the mismatch minimum around the mass-spin plane is reduced as well, however it still doesn't settle on the location of the true remnant properties.
This behaviour may explain some of the observed widening of the histograms, and perhaps hints something is missing from the ringdown model.

\subsection{Higher Harmonics}\label{kitchen-sink}

As demonstrated by Fig.~\ref{misaligned_spin_epsilon_hist} (and also Fig.~\ref{misaligned_spin_epsilon_hist_m10} in appendix \ref{appendix_a}), the overtone and mirror mode models considered so far achieve median values for the remnant error $\epsilon \sim 0.01$, a factor of 10 or more higher than the aligned-spin fits of Fig.~\ref{aligned_spin_epsilon_hist}. In addition, the spread of $\epsilon$ values recovered is significantly larger, leading to values of $\epsilon$ up to $\sim 0.1$. 
These models perform significantly worse in some cases for precessing systems than aligned-spin systems.

We now investigate whether the inclusion of higher harmonics (that is, QNMs with $\ell > 2$) can improve the fits to $h'_{22}(t)$.
These higher harmonics were neglected by both the overtone (Eq.~\ref{GieslerRD}) and mirror mode (Eq.~\ref{mirror_ringdown}) models.
However, mode mixing does occur as a consequence of the different angular basis functions used in the waveform decompositions in Eqs.~\ref{YlmExpansion} and \ref{general_ringdown} and the fact that these basis functions are not mutually orthogonal \cite{2014PhRvD..90f4012B}.
The amount of mode mixing between the spherical mode ${}_{-2}Y_{\ell m}$ and the spheroidal mode ${}_{-2}S_{\ell m n}$ is determined by the remnant spin $\chi_f$ and the QNM frequency. This can be quantified by how much these functions fail to be orthogonal; i.e.\ by the integral
\begin{align}
     \mu_{\ell m, \ell' m' n'} = \delta_{mm'}\int_{\Omega} {}_{-2}Y_{\ell m}(\Omega)  ~ {}_{-2}S^*_{\ell ' m' n'}(\Omega) ~ \dd{\Omega},
\end{align}
where $\Omega$ denotes the angles $\theta,\,\phi$.
A translational offset between the NR and ringdown frames (e.g.\ due to a kick) can also lead to mixing between $m$-modes \cite{2016PhRvD..93h4031B}; this effect is neglected here.
To include the contribution from higher harmonics, we define a new ringdown model for the spherical harmonic modes which now allows for a sum over different $\ell$:
\begin{align}\label{full_ringdown}
    h_{\ell m}^{N,\,L,\, {\rm mirror}}(t) = &\sum_{n=0}^N \sum_{l=2}^{L} \Big[ C_{l m n} e^{-i \omega_{l m n}(t-t_0)} \\ &+ C'_{l m n} e^{i \omega^*_{l m n}(t-t_0)} \Big]\quad \textrm{for} \quad t \geq t_0. \nonumber
\end{align}
This \emph{harmonic} model contains all of the allowed QNMs in Eq.~\ref{general_ringdown}, including the mirror modes and the overtones.
This comes at the expense of a large number of free parameters; there are $4(N+1)(L-\ell+1)$ in the complex amplitudes, plus the two remnant parameters $M_f,\; \chi_f$ that determine the complex QNM frequencies.

Multiple variations of this harmonic model were trialled (varying $N$, $L$, and the inclusion of mirror modes) on the same population of 252 misaligned-spin SXS simulations.
Fig.~\ref{misaligned_spin_epsilon_hist_harmonics} shows the chosen subset of results, which includes the $N=7$, $L=3$ and $L=4$ models (both with mirror modes). As before, we fit to the rotated $h'_{22}(t)$ spherical harmonic mode.
To make a clear comparison with the previous models, we again use a ringdown start time corresponding to the peak of the GW energy flux.

The inclusion of higher harmonics (dashed histograms in Fig.~\ref{misaligned_spin_epsilon_hist_harmonics}) drastically improves the mismatch.
A small mismatch is not surprising for a model with so many free parameters, and in some of these cases we are likely pushing beyond the limits of accuracy of the NR simulations. See appendix \ref{NR_error_appendix} for a discussion of the numerical errors.
There is a modest reduction in $\epsilon$ for some systems, and in particular we see less systems with $\epsilon > 0.01$ (at least for $L=3$). This hints at the importance of higher harmonics in some precessing systems. Despite this, we still see worst-case values of $\epsilon \sim 0.04$.

\section{Surrogates}\label{surrogate-section}

NR simulations are computationally expensive, and although the number of simulations available in public catalogs is growing they are still limited in their parameter space coverage. 
NR surrogate models \cite{2015PhRvL.115l1102B, 2015PhRvL.115l1102B, 2017PhRvD..96b4058B, 2019PhRvR...1c3015V, 2019PhRvD..99f4045V} would appear to be an attractive alternative.
These models use reduced-order and surrogate modelling techniques to extend the results of a set of NR simulations smoothly across parameter space. 
The use of surrogates could, in principle, allow us to extend the results of the previous section to include many more systems as well as allowing us to study how the excitations of the various QNMs vary during a smooth exploration of parameter space.
However, care must be taken as the surrogate modelling necessarily introduces an additional source of error into the waveforms, on top of the errors originally in the NR waveforms themselves.
 
When attempting to fit QNM ringdown models with overtones to NRSur7dq4 \cite{2019PhRvR...1c3015V} waveforms, it was found that incorrect values for $M_f$ and $\chi_f$ were being recovered (particularly at high mass ratios). This being the case even for aligned-spin or non-spinning systems. Although the NRSur7dq4 waveforms do not provide the remnant properties, these can be obtained via NRSur7dq4Remnant \cite{2019PhRvR...1c3015V} (it was found the problem did not lie with the values returned by NRSur7dq4Remnant but rather with the waveform surrogate).

\begin{figure}[b]
    \centering
    \includegraphics[width=\columnwidth]{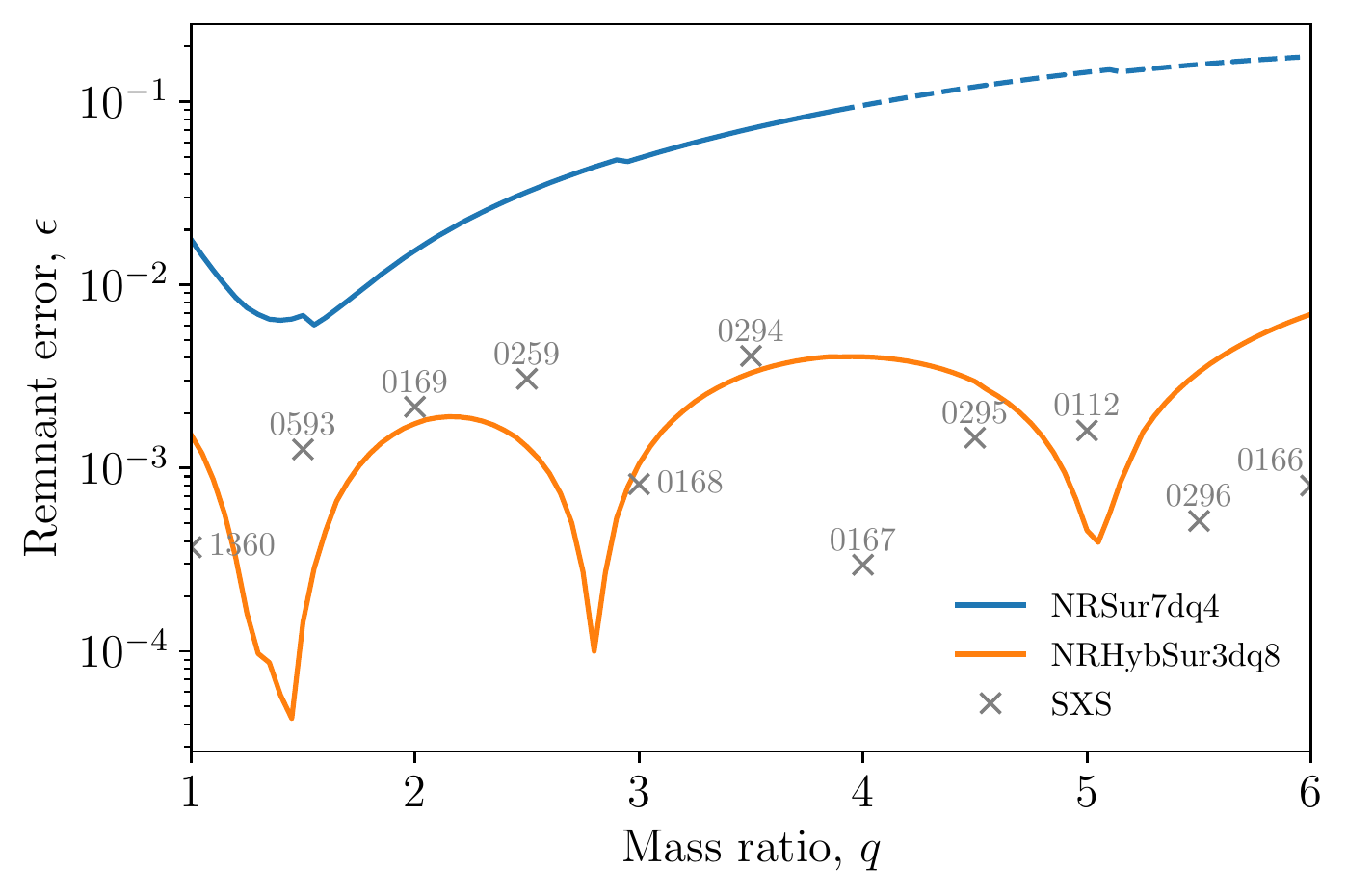}
    \caption{
    Comparison of the remnant error $\epsilon$ from two surrogate models and a selection of SXS simulations. All are zero initial spin. The fits were performed on the $h_{22}$ mode with the $N=7$ overtone model, Eq.~\ref{GieslerRD}, starting from the time of peak strain. The labels on each cross correspond to the SXS ID. The dashed line indicates where we are outside the training range of NRSur7dq4.
    }
    \label{surrogate_epsilon_vs_q}
\end{figure} 

To investigate the performance of NRSur7dq4 ringdown waveforms, a series of simulations with zero initial spin with increasing mass ratio $q$ from 1 to 6 were used. 
The $N=7$ overtone model (Eq.~\ref{GieslerRD}) was fitted to the $h_{22}(t)$ mode of each starting from the peak strain (as in section \ref{aligned-spin-section})
and the remnant error $\epsilon$ (Eq.~\ref{eq:epsilon}) was calculated for each.
The results are shown in Fig.~\ref{surrogate_epsilon_vs_q}, along with the results for similar fits performed directly on 11 zero-spin SXS simulations at discrete values of the mass ratio. 
The fits to the NRSur7dq4 surrogate produce values for $\epsilon$ that are 1-2 orders of magnitude higher than for the equivalent SXS simulations. 
Also shown are the results from a similar analysis with the more restrictive aligned-spin surrogate NRHybSur3dq8 \cite{2019PhRvD..99f4045V}; this was found to be in close agreement with the SXS simulations.

Residuals and mismatches can also be computed between surrogate and NR waveforms (taking care to align the waveforms in both time and phase).
For SXS:BBH:0168, the $q=3$, zero-spin simulation used in Fig.~\ref{surrogate_epsilon_vs_q}, we find $\sim 2\%$ residuals in the ringdown when comparing to the NRSur7dq4 surrogate with the same parameters. 
This leads to a mismatch between the surrogate and SXS:BBH:0168 of $3.7 \times 10^{-4}$, when integrating over the ringdown. For comparison, we have a $\sim 10^{-6}$ mismatch between the ringdown model Eq.~\eqref{GieslerRD} and the SXS simulation. The relatively high mismatch between the NRSur7dq4 and SXS waveforms translates to the relatively high values of $\epsilon$ seen in Fig.~\ref{surrogate_epsilon_vs_q}. 

It seems that the high-dimensional precessing surrogate NRsur7dq4 is not yet sufficiently accurate in the ringdown for the purposes of QNM overtone studies that, by virtue of their large number of free parameters, fit the ringdown with very small mismatches. 
By contrast, the lower-dimensional aligned-spin surrogate NRHybSur3dq8 does appear to be sufficiently accurate for such studies.

\section{Discussion} \label{sec:discussion}

This paper has made a first systematic attempt at using QNMs to model the ringdown of BHs formed from BBHs with misaligned component spins in the inspiral.
Previously, for aligned-spin systems, it has been found that the ringdown can be modelled with low mismatch and low remnant errors using a model that includes overtones of the fundamental QNM \cite{overtones}. For seven overtones, the ringdown can be reliably modelled from the peak of the $h_{22}(t)$ strain for a range of SXS simulations.
Additionally, the inclusion of mirror modes can allow the ringdown to be modelled from even earlier times \cite{mirror_modes}.
In this paper, which generalised these studies to precessing systems, we find that while QNM models can reliably achieve small mismatches, in the worst cases the remnant errors are more than a factor of 10 higher.
This is the case even when choosing to start the ringdown at the more conservative (i.e.\ later) peak in GW energy flux. 
The inclusion of higher harmonics reduces the remnant error in some cases, perhaps a sign that mode mixing in the ringdown is generally more important in precessing systems. However, in other cases, a bias remains in the recovered remnant properties.
We conclude that it is not possible to reliably model the ringdown from the peak in the flux, or indeed from the peak in the strain. 

We end by sounding a brief note of caution to any who attempt to construct a QNM model starting at or before the peak flux or strain. 
While such a model will work in some cases, it risks biased results in others. 
This risk is subtle because QNM models can give small mismatches even when they fail to adequately describe the remnant.


\begin{acknowledgments}
    We thank Alberto Vecchio, Davide Gerosa, Vijay Varma, Geraint Pratten, Nathan Johnson-McDaniel, Gregorio Carullo and Matthew Giesler for helpful discussions and comments.
    This document has been assigned LIGO document number P2100036.
    Some computations were performed using the University of Birmingham's BlueBEAR HPC service.
\end{acknowledgments}


\bibliographystyle{apsrev4-1}
\bibliography{references}


\clearpage

\onecolumngrid

\appendix

\section{Overtone Model Fits to a Variety of Precessing NR Simulations}\label{appendix_a}

\begin{figure}[h]
    \centering
    \includegraphics[width=1\textwidth]{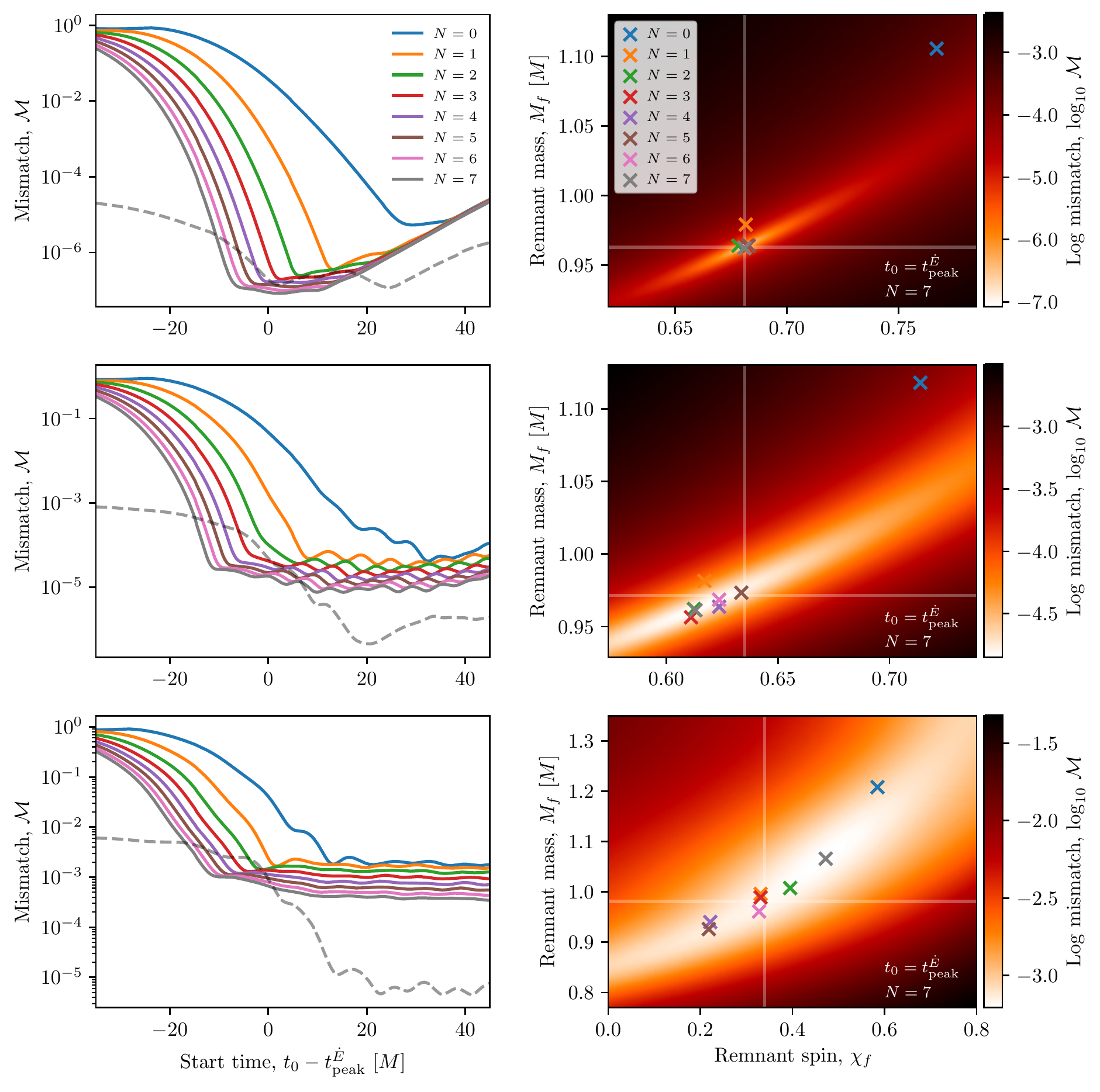}
    \caption{ \label{misaligned_spin_variation}
    A selection of results for modelling the ringdown of precessing NR simulations from the SXS catalog \cite{sxs, 2013PhRvL.111x1104M,sxs_catalog} using the overtone model in Eq.~\ref{GieslerRD}.
    These plots show the results for the three systems described in the table that have been chosen to illustrate the wider range of behaviours that occur for precessing systems, from good at the top to bad at the bottom.
    The left-hand column of plots also shows the difficulty in identifying a general start time for the ringdown as mismatch is minimised for a range of different times and sometimes there isn't even a clear first minimum.
    }
\end{figure}

\begin{footnotesize}
\begin{center}
\begin{tabular}{ c|c|c|c|c|c } 
$\;$SXS:BBH ID $\;$ & $\;$Figure row$\;$ & $\;$Remnant error $\epsilon$$\;$ ($\epsilon_{\mathrm{NR}}$) &  $\;$Mass ratio $q$$\;$ & Component spins $\vb*{\chi}_1$, $\vb*{\chi}_2$  & $\;$Remnant spin $\vb*{\chi}_f$$\;$ \\
\hline
1677 & top & $8.1 \cross 10^{-4}$ ($1.8 \cross 10^{-4}$) & 2.64 & $(-0.06,\,0,\,0.27)$, $(-0.49,\,-0.55,\,0.06)$ & $(-0.05,\,0,\,0.68)$ \\ 
1768 & middle & $2.6 \cross 10^{-2}$ ($8.0 \cross 10^{-4}$) & 3.49 & $(0.65,\,0.03,\,0.01)$, $(-0.3,\,0.05,\,0.47)$ & $(0.31,\,-0.02,\,0.56)$ \\ 
1789 & bottom & $1.6 \cross 10^{-1}$ ($4.8 \cross 10^{-4}$) & 3.72 & $(0.46,\,0.08,\,-0.52)$, $(-0.43,\,-0.28,\,-0.17)$ & $(0.14,\,0.01,\,0.31)$ \\ 
\end{tabular}
\end{center}
\end{footnotesize}

\section{Overtone Model Fits to a Population of Precessing NR Systems Starting Before the Peak Flux}\label{misaligned_spin_fits_appendix}

\begin{figure*}[h]
    \centering
    \includegraphics[width=\textwidth]{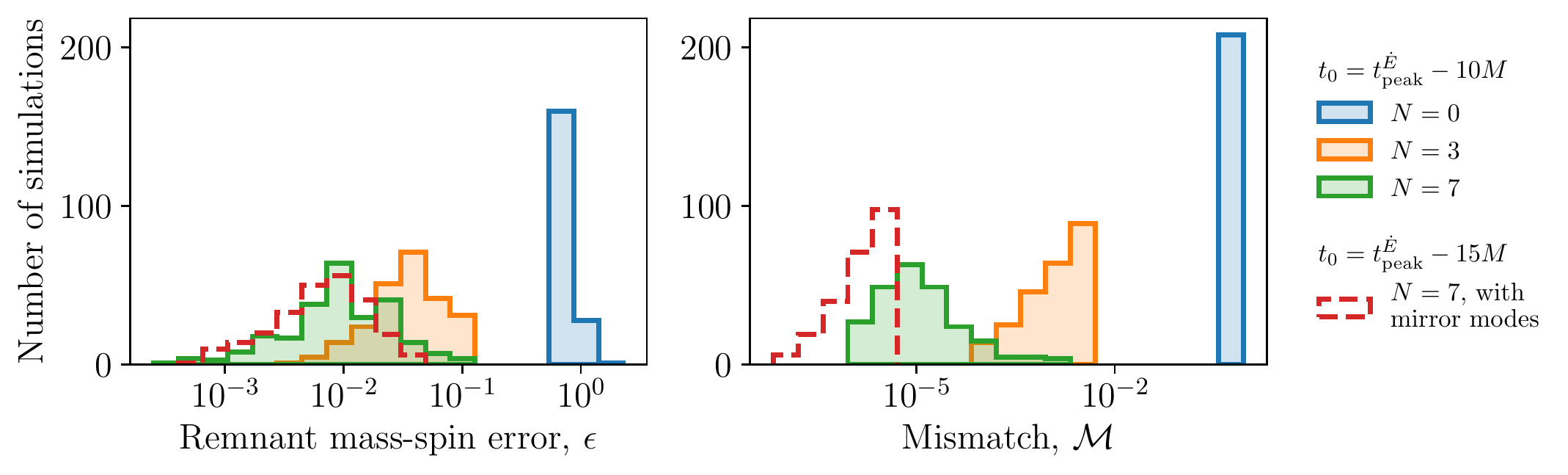}
    \caption{
    Left: histograms of the mass-spin remnant error $\epsilon$ from an overtone model fit to the rotated $h'_{22}$ mode of 252 misaligned-spin SXS simulations for several different overtone numbers $N$. 
    Right: histograms of the mismatch from a fit with the true remnant mass and spin parameters, with the same overtone models and SXS simulations as in the left histogram.
    These results are similar to those in Fig.~\ref{misaligned_spin_epsilon_hist} in the main text, but use a start time that is earlier by $10M$.
    The solid histograms show results from fits performed starting $10M$ before the peak of the energy flux with $N$ overtones of the fundamental $\ell = m = 2$ mode.
    The red dashed line shows results from a $N=7$ model that also includes mirror modes and was fitted with a ringdown starting $15M$ before the peak in the energy flux.
    }
    \label{misaligned_spin_epsilon_hist_m10}
\end{figure*}

The analysis on the population of misaligned-spin simulations performed in section \ref{misaligned-spin-section} (results plotted in Fig.~\ref{misaligned_spin_epsilon_hist}) is repeated here using an earlier start time for the ringdown: $t_0=t^{\dot{E}}_{\rm peak}-10M$.
This was done to check whether a poor choice of start time was responsible for some of the poor fits obtained using the overtone model in Eq.~\ref{GieslerRD}.
The new results are plotted in Fig.~\ref{misaligned_spin_epsilon_hist_m10}.
We find that the $N=7$ model results do not significantly change with the new start time.
The $N=3$ and $N=0$ model results do change and generally give a worse fit with the earlier start time, as might be expected. This analysis shows that the overtone model (with or without mirror modes) cannot be reliably applied to precessing systems at early times. 

\section{Numerical Relativity Errors}\label{NR_error_appendix}

It is important to remember the finite accuracy of the NR simulations used in ringdown studies.
This is particularly true when using models with many QNMs which, by their very nature, use a large number of free parameters and regularly achieve very small ($\sim 10^{-6}$) mismatches.
If care is not taken, we risk fitting our models to the numerical noise. 
In this appendix we describe the numerical checks performed on the 5 individual simulations used in this paper: SXS:BBH:0305, 1856, and the three simulations shown in Fig.~\ref{misaligned_spin_variation}.
In each case the numerical errors were estimated by comparing results obtained using data from the two highest resolutions (levels) available in the SXS catalog. 

First, we quantify the numerical error in the mismatch.
This was done by calculating the mismatch between the two NR resolutions from a time $t_0$ to a time $T = t_0 + 100M$, for a range of $t_0$. For each start time, we optimally align the two waveforms in time (taking the absolute value in the mismatch automatically optimises the mismatch over phase). The alignment in time can be done by matching the time of peak strain, for example, or by numerically rolling the waveform to find the optimal time shift for each mismatch calculation.
The results are shown by the grey dashed lines in the mismatch vs start time plots in Figs.~\ref{305_mismatch_vs_t0}, \ref{1856_mismatch_vs_t0} (duplicated in Fig.~\ref{1856_mirror_mode_mismatch_vs_t0}) and the 3 panels of Fig.~\ref{misaligned_spin_variation}.
Generally, we see numerical error estimates at or below the model mismatches, particularly at late times, indicating that we are not fitting to the numerical noise.
The main exception is Fig.~\ref{1856_mirror_mode_mismatch_vs_t0} where the mirror mode model is applied to a precessing system. This is expected; precessing NR simulations, and those with high mass ratios are generally expected to have larger numerical errors. Additionally, the mirror mode and harmonic models have the highest numbers of free parameters making them more likely to reach the accuracy of the NR simulation. 

Second, we investigate the numerical error on the remnant mass and spin.
We quantify the numerical error with $\epsilon_{\mathrm{NR}}$, the Euclidean distance (Eq.~\ref{eq:epsilon}) between the remnant properties reported in the two highest resolution levels of the NR simulation.
The $\epsilon_{\mathrm{NR}}$ values are reported in the main text and in the table in appendix \ref{appendix_a}.
In all cases $\epsilon_{\mathrm{NR}} < \epsilon$. This supports the conclusions in the main text and indicates they are likely to be robust against numerical noise in the underlying NR simulations used.

\end{document}